\documentclass[twocolumn,10pt]{revtex4-2}
\usepackage{amsmath}

\usepackage{amssymb}

\newcommand{\be}{\begin{equation}}
\newcommand{\ee}{\end{equation}}
\newcommand{\bea}{\begin{eqnarray}}
\newcommand{\eea}{\end{eqnarray}}
\begin{document}

\title{Migdal-Eliashberg and SUS-$Y^2$-SYK}

\author{D. V. Khveshchenko}
\affiliation{Department of Physics and Astronomy, 
University of North Carolina, Chapel Hill, NC 27599}

\begin{abstract}
\noindent
This article addresses a number of subtle issues pertaining to the long-standing problem of strong phonon-like fermion-boson coupling. Among the central topics are the customary Migdal-Eliashberg approximation in the pertinent Schwinger-Dyson gap equation and its solutions. The previously gained insight is assessed by contrasting it against the various (non-)supersymmetric variants of 
the Yukawa-Sachdev-Ye-Kitaev model. 
Also, some previously discussed (pseudo-)holographic aspects of fermion pairing in such models are commented upon.  
\end{abstract}
\maketitle

\noindent
{\it Flat band in flatland}

\vspace{\baselineskip}

\noindent
Cross-fertilization between string/high-energy and condensed matter physics has brought out
many constructions that routinely would be studied separately and without seeking out a connection. 
Finding such links may turn out to be highly beneficial for both, though.
 
For one, the problem of (nearly) dispersion-less electrons coupled to optical phonons has a rich history.
The central issue at stake is the competition between a possible non-Fermi liquid (NFL) behavior 
and onset of superconductivity. 
A better understanding of the phenomenon of electron pairing in the absence of well-defined quasiparticles could provide additional guidance in searching for better and more robust superconductors. To that end, it is important to find out whether an observation of the notorious 'strange metallic' NFL features in the normal state could be a precursor of potentially higher critical temperature, magnetic field, current, etc. 
In that regard, especially promising are low-dimensional systems ('flatlands') with their strong interaction effects,   
subsequent loss of quasiparticle coherence, and ensuing alternatives to the conventional (quasiparticle-based) 
Bardeen-Cooper-Schrieffer (BCS) pairing scenario.
 
The continuing quest into solvable examples of the NFL systems and their intrinsic instabilities  
brought out, first, the extensively studied Sachdev-Ye-Kitaev model (SYK)
and, subsequently, its further Yukawa extension (YSYK).
Originally, the former was introduced to study spin glasses (or, still earlier, complex nuclei) \cite{french_random-matrix_1971,bohigas_spacing_1971,sachdev_gapless_1993}, 
although later it was refurbished as a model of rigid clusters of dispersionless ('flat-band')
Majorana or Dirac fermions governed by the strongly disordered, yet distance-independent, all-to-all interactions. 
Beyond this specific context it was 
also portrayed as a waterproof example of low-dimensional holography \cite{Kitaev2015,kitaev_notes_2018,kitaev_soft_2018,kitaev_statistical_2019,sachdev_holographic_2010,sachdev_bekenstein-hawking_2015,sachdev_strange_2023,maldacena_bound_2016,maldacena_remarks_2016,maldacena_conformal_2016,stanford_fermionic_2017,polchinski_spectrum_2016,gross_generalization_2017,gross_all_2017,sarosi_ads_2_2018,bagrets_sachdev-ye-kitaev_2016,bagrets_power-law_2017,mertens_solving_2017,mertens_schwarzian_2018,yang_quantum_2019,jha_theory_2026,khveshchenko_thickening_2018,khveshchenko_seeking_2018,khveshchenko_true_2019,khveshchenko_one_2020,khveshchenko_gloria_2022}. 

Subsequently, the YSYK model was proposed as the description of a (disordered) elastic 
quantum dot harboring $N$ localized electron orbitals which are coupled via $M$ optical phonon modes \cite{chubukov_superconductivity_2005,bi_instability_2017,classen_superconductivity_2021,wang_phase_2021,hauck_eliashberg_2020,wang_solvable_2020,esterlis_cooper_2019,chowdhury_intrinsic_2020,pan_yukawa-syk_2021,aldape_solvable_2022,choi_pairing_2022,chowdhury_sachdev-ye-kitaev_2022,wang_quantum_2020,bashan_tunable_2024,li_strange_2024,davis_quantum_2023,valentinis_correlation_2023,valentinis_bcs_2023,lantagne-hurtubise_superconducting_2021,zhang_density_2023,elezaby_superconductivity_2026,gnezdilov_upper_2025,sutradhar_singlet_2024,bonetti_fractionalized_2026,kleger_universal_2026}. 
Built-in randomness of the coupling parameters in both models is often considered to be crucially 
important for developing the NFL behaviors that have been found to agree with the experimental 
observations in the cuprates and other documented 'strange metals'.    

From the formal standpoint, random ensemble averaging greatly simplifies matters by 
selecting the so-called 'melonic' graphs as the dominant class of Feynman diagrams in the large-$N$ limit \cite{klebanov_uncolored_2017,peng_vector_2017,fraser-taliente_melonic_2024}.
In the condensed matter context, this selection 
corresponds to the celebrated Migdal-Eliashberg (ME) approximation 
in the underlying Schwinger-Dyson (SD) equations 
which systematically neglects vertex corrections \cite{migdal_interaction_1958,eliashberg_interactions_1960,eliashberg_temperature_1963,alexandrov_breakdown_2001,marsiglio_eliashberg_2020,bauer_quantitative_2011,chowdhury_unreasonable_2020,yang_non-perturbative_2023,pan_nonperturbative_2021,huang_theoretical_2023,huang_superconductivity_2025,schrodi_full-bandwidth_2020,damia_thermal_2021,wang_non-fermi_2017,wang_superconductivity_2018}.
The ME approximation improves 
on the basic BCS theory by incorporating possible electron mass renormalization and  
phonon retardation effects due to the electron polarization.

It is worth noting, though, that the previously found similarity between the properties of 
the original random SYK model and its non-random tensor counterparts \cite{witten_syk-like_2016,gurau_complete_2017,gurau_1n_2011} suggest that disorder averaging in SYK 
may not be that important for achieving melonic dominance, after all.
\\

\noindent
{\it {SD equations}}
\\

\noindent   
In the system of fermions with a momentum-dependent dispersion 
$\xi_{\bf k}$ 
which are coupled to bosons with a propagator $D_0(\omega,{\bf k})$,    
the matrix-valued (Nambu) fermion self-energy is comprised of the 
normal-state $\Sigma(\omega,{\bf k})$ and pairing  $\Phi(\omega,{\bf k})$ components 
\be
{\hat {\varSigma}}(\omega,{\bf k})=
\begin{pmatrix}
\Sigma(\omega,{\bf k}) & \Phi(\omega,{\bf k})\\
\Phi^{*}(-\omega,-{\bf k}) & -\Sigma(-\omega,-{\bf k})
\end{pmatrix}
\label{eq:nambu-sigma}
\ee
Together with the boson polarization $\Pi(\omega,{\bf k})$ it obeys the standard SD equations 
\cite{migdal_interaction_1958,eliashberg_interactions_1960,eliashberg_temperature_1963,alexandrov_breakdown_2001,marsiglio_eliashberg_2020,bauer_quantitative_2011,chowdhury_unreasonable_2020,yang_non-perturbative_2023,pan_nonperturbative_2021,huang_theoretical_2023,huang_superconductivity_2025,schrodi_full-bandwidth_2020,damia_thermal_2021,wang_non-fermi_2017,wang_superconductivity_2018}
\bea
{\hat {\varSigma}}(\omega,{\bf k})={\hat {\cal G}}^{-1}(\omega,{\bf k})-{\hat {\cal G}}_0^{-1}(\omega,{\bf k})= \nonumber\\
\int_{\omega^{\prime},k^{\prime}}
\Lambda(\omega,\omega^{\prime};{\bf k}, {\bf k}^{\prime}){\hat {\cal G}}(\omega^{\prime},{\bf k}^{\prime})
D(\omega-\omega^{\prime},{\bf k}-{\bf k}^{\prime})
\label{eq:sd-fermion}
\eea
and
\bea
\Pi(\omega,{\bf k})=D^{-1}(\omega,{\bf k})-D_0^{-1}(\omega,{\bf k})=\nonumber\\
\int_{\omega^{\prime},{\bf k}^{\prime}}
Tr{\hat {\cal G}}(\omega+\omega^{\prime},{\bf k}+{\bf k}^{\prime})
\Lambda(\omega,\omega^{\prime};{\bf k}, {\bf k}^{\prime})
{\hat {\cal G}}(\omega^{\prime},{\bf k}^{\prime})
\label{eq:sd-boson}
\eea
where the (inverse) fermion propagator 
\be
{\cal {\hat G}}^{-1}(\omega,{\bf k}) = 
\begin{pmatrix}
i\omega-\xi_{\bf k} & 0\\
0 & i\omega+\xi_{\bf k}
\end{pmatrix}
+{\hat \varSigma}(\omega,{\bf k})
\label{eq:ginv-nambu}
\ee
incorporates the matrix-valued 
self-energy 
\eqref{eq:nambu-sigma} 
and consists of the normal ($G)$ and anomalous ($F$) functions 
\be 
{\hat {\cal G}}(\omega,{\bf k})=
\begin{pmatrix}
G(\omega,{\bf k}) & F(\omega,{\bf k})\\
F^{*}(-\omega,-{\bf k}) & -G(-\omega,-{\bf k})
\end{pmatrix}
\label{eq:g-nambu}
\ee
whereas $\Lambda(\omega,\omega^{\prime};{\bf k},{\bf k}^{\prime})$ in Eqs.(2,3) is the vertex function.

The free (condensation) energy difference between the 
normal and paired states is computed as \cite{tsoncheva_condensation_2005,marsiglio_electron_2001,mirabi_thermodynamics_2020,zhang_superconductivity_2022,zhang_free_2023-1}
\bea
\delta F=\frac{1}{2}\sum_{{\bf k},{\omega};{\bf k^{\prime}},\omega^{\prime}}
Tr{\hat \Sigma}(\omega,{\bf k})
\Lambda(\omega,\omega^{\prime};{\bf k},{\bf k}^{\prime})
{\hat \Sigma}(\omega^{\prime},{\bf k}^{\prime})
\nonumber\\
-T\sum_{{\bf k},{\omega}}{\sqrt {(i\omega-\xi_{\bf k}+\Sigma(\omega,{\bf k}))^2+|\Phi(\omega,{\bf k})|^2}}~~~~~~~~~
\nonumber\\
\eea
In the case of multiple solutions to Eqs.(1-5), comparing their free energies provides the way of identifying the most stable one - hence, the first (or mostly likely) one to develop.  

The previous studies of the electron-phonon systems  
revealed a spurious instability 
at moderate (bare) electron-phonon couplings, at which point the effective coupling 
diverges while the phonon spectrum flattens out.
This behavior was identified as possibly signaling 
either an emergent charge density wave (CDW) or (bi-)polaron formation \cite{paci_polaronic_2005,fratini_optical_2006,heidarian_inhomogeneous_2003,wolfle_self-consistent_2010,miskic_polaron_2025,ragni_polarons_2025,chubukov_breakdown_2026,prokofev_limits_2026,chubukov_eliashberg_2020,cai_superconductivity_2025,kabanov_adiabatic_2020}.

Furthermore, it was found that at strong electron-phonon couplings the electronic specific heat becomes negative  \cite{yuzbashyan_superconductivity_2022,yuzbashyan_breakdown_2022,yuzbashyan_migdal-eliashberg_2022,yuzbashyan_instability_2025,kiessling_bounds_2025,kiessling_bounds_2025-1,kiessling_bounds_2025-2,
semenok_fundamental_2025}.  
Despite the fact that the overall specific heat remains positive due to the (much larger) phonon contribution, it was conjectured that the negative electronic specific heat signals a potential kinetic instability and 
intrinsically non-equilibrium behavior of the coupled electron-phonon system.
Moreover, recently it was argued that the kinetic instability may set in at even weaker couplings and before the electron specific heat changes sign \cite{yuzbashyan_superconductivity_2022,yuzbashyan_breakdown_2022,yuzbashyan_migdal-eliashberg_2022,yuzbashyan_instability_2025,kiessling_bounds_2025,kiessling_bounds_2025-1,kiessling_bounds_2025-2,semenok_fundamental_2025}.
\\

\noindent
{\it ME approximation}
\\

\noindent
Depending on the nature of the fermions' interaction the kernel $D_0(\omega,{\bf k})$ in Eqs.(1-6)
can be strongly localized 
in the position space and rapidly (albeit not instantaneously) decaying in the time domain 
($D_0(\omega,{\bf k})\sim 1/(\omega^2+\Omega^2)$, 
as for optical phonons), conform to the spectral density of
linear acoustic phonons of speed $v_s$ ($D_0(\omega,{\bf k})\sim 1/(\omega^2+v_s^2{\bf k}^2)$, 
or take a more general 'non-Ohmic' form ($D_0(\omega,{\bf k})\sim 1/(|\omega|^{\gamma}+v_s^2{\bf k}^2)$ 
with $\gamma\neq 2$), as in the quantum-critical (potentially, NFL) regime where
such coupling can be generated by the polarization of the fermions themselves.

Under the conditions of applicability of the ME approximation a typical value of the ratio 
$\omega/\xi_k$ in phonon emission/absorption is of order $v_s/v_F\ll 1$ for acoustic phonons
and $\omega_D/\epsilon_F\ll 1$ (except for the extreme low-density systems) for their optical counterparts.
When focusing on the energy dependence it is then customary to assume the self-energy and pairing function to be
(nearly) independent of the absolute value of momentum, except for a possible 
angular dependence.

For fermions with a non-flat spatial dispersion and an extended Fermi surface, applying the ME argument 
about the smallness of vertex corrections, $\Lambda(\omega,\omega^{\prime};{\bf k}, {\bf k}^{\prime})\approx g=const$, 
(hence, approximate independence of the self-energy matrix ${\hat {\varSigma}}$ of momentum),  
facilitates the momentum integration in Eqs.(2,3). 

From the technical standpoint, momentum independence is more important than any remaining energy dependence. 
Therefore, in what follows we will consider a more general situation by allowing the latter. 
One then finds that the (renormalized) electron dispersion appears only in the denominators of Eqs.(2,3), 
thus enabling straightforward momentum integration.
Introducing the Fermi surface density of states (DOS) $\nu_F$,
one then arrives at the coupled equations for the energy-dependent functions   
\be
\Sigma(\omega)=\nu_F\int_{\omega^{\prime}}\Lambda(\omega,\omega^{\prime})
d(\omega-\omega^{\prime}){i\omega^{\prime}+\Sigma(\omega^{\prime})\over \Delta(\omega^{\prime})}
\label{eq:me-sigma}
\ee
and 
\be
\Phi(\omega)=\nu_F\int_{\omega^{\prime}}\Lambda(\omega,\omega^{\prime})
d(\omega-\omega^{\prime}){\Phi(\omega^{\prime})\over \Delta(\omega^{\prime})}
\label{eq:me-phi}
\ee
where $\Delta(\omega)= {\sqrt {(i\omega+\Sigma(\omega))^2+|\Phi(\omega)|^2}}$ and 
the kernel is given by the interaction function $D(\omega,{\bf k})$ averaged over the Fermi surface
\be
d(\omega)=<{D_0(\omega,{\bf k})
\over {1+g\Pi(\omega,{\bf k})D_0(\omega,{\bf k})}}>_{FS}
\label{eq:kernel-fs}
\ee 
Notably, the powers of 
$\Delta(\omega)$ in the denominators of Eqs.(7,8) are reduced by one, as compared to Eqs.(2,3).
\\

\noindent
{\it Equivalent spin chain}
\\

\noindent 
Solutions to the SD equations (7,8) can be elegantly 
described as the minimal energy configurations of a $1d$ chain of (normalized) classical spins \cite{yuzbashyan_superconductivity_2022,yuzbashyan_breakdown_2022,yuzbashyan_migdal-eliashberg_2022,yuzbashyan_instability_2025,kiessling_bounds_2025,kiessling_bounds_2025-1,kiessling_bounds_2025-2,semenok_fundamental_2025} 
\be
{\bf S}_n={1\over \Delta(\omega_n)}[i\omega_n+\Sigma(\omega_n), Re\Phi(\omega_n), Im\Phi(\omega_n)]
\label{eq:spin-map}
\ee
labeled by the site number $n$ which corresponds to the Matsubara frequency $\omega_n=2\pi T(n+1/2)$.

The effective spin-chain Hamiltonian 
\be
H=\sum_n{\bf B}_n{\bf S}_n-{1\over 2}\sum_{n,m} J_{nm} {\bf S}_n{\bf S}_m
\label{eq:spin-hamiltonian}
\ee
is equivalent to the free energy (6), featuring the site-dependent magnetic field ${\bf B}_n={\bf \hat z}\omega_n$ and long-range ferromagnetic exchange 
coupling $J_{nm}=d(\omega_n-\omega_m)$.
In equilibrium, the equation of motion stemming from (11) reads
\be
{\bf S}_n\times{\bf B}_n-\sum_m J_{nm}{\bf S}_n\times{\bf S_m}=0
\label{eq:spin-torque}
\ee
Alternatively, the frequency $\omega_n$
can be thought of as a discretized vertical coordinate, the first term in (11) then representing gravitational potential energy. 

To extend this description to the non-stationary and/or non-equilibrium situations 
one can utilize the coherent state representation spanned by the (overcomplete)  
basis of eigenstates, ${\hat {\bf S}}|{\bf n}>={\bf n}|{\bf n}>$.  
In the continuum (low-temperature) limit the real-time dynamics of the unit vector ${\bf n}(\tau)$ 
is described by the $1+0$-dimensional 
non-linear $\sigma$-model on the 
coset $SO(3)/SO(2)$, 
parametrized by the spherical 
angles $\phi$ and $\theta$ 
\bea
S={1\over 2}\int_{\tau}\cos\theta(\tau)\partial_{\tau}\phi(\tau)\nonumber\\
-\int_{\tau}
({\bf B}(\tau){\bf n}(\tau)+J(\tau){\bf n}(\tau){\bf n}(-\tau)))~~~~~~~~~\
\label{eq:sigma-model}
\eea
The equation of motion derived from the action (13)
conforms naturally to the Landau- Lifshitz one  
\be
{\bf n}(\tau)\times\partial_{\tau}{\bf n}(\tau)-({\bf B}(\tau)+J({\tau}){\bf n}(-{\tau}))=0
\label{eq:landau-lifshitz}
\ee
An alternate approach to the SD equations in the ME approximation employs the so-called Anderson spins \cite{anderson_new_1958,anderson_random-phase_1958}
which emerge from integrating Eqs.(2,3) over the frequency $\omega$ instead of the fermion dispersion $\xi_{\bf k}$.
\\

\noindent
{\it Ultra-local limit}
\\

\noindent
In the flat-band limit of $\xi_{\bf k}\to const$, 
the aforementioned 'radial' integration in the momentum space becomes trivial, resulting in one
greater power of the function $\Delta(\omega)$ in the denominators of Eqs.(7,8). 
Correspondingly, the effective spin representation does not naturally emerge.
Instead, the coupled equations (2,3) take the form
\be
\Sigma(\omega)=\int_{\omega^{\prime}}
\Lambda(\omega,\omega^{\prime})d(\omega-\omega^{\prime})
{i\omega^{\prime}+\Sigma(\omega^{\prime})\over \Delta^2(\omega^{\prime})}
\label{eq:ul-sigma}
\ee
and 
\be
\Phi(\omega)=\int_{\omega^{\prime}}
\Lambda(\omega,\omega^{\prime})d(\omega-\omega^{\prime})
{\Phi(\omega^{\prime})\over \Delta^2(\omega^{\prime})}
\label{eq:ul-phi}
\ee
where the fermion polarization reads  
\bea
\Pi(\omega)=\int_{\omega^{\prime}}
\Lambda(\omega+\omega^{\prime},\omega^{\prime})\times~~~~~~~~~~~\\
{(i\omega^{\prime}+\Sigma(\omega^{\prime}))
(i\omega+i\omega^{\prime}+\Sigma(\omega+\omega^{\prime}))
-\Phi^{*}(\omega^{\prime})\Phi(\omega+\omega^{\prime})
\over {\Delta^2(\omega+\omega^{\prime}) \Delta^2(\omega^{\prime})}}\nonumber
\label{eq:ul-pi}
\eea
Furthermore, in the limit of extremely strong coupling ($g\to\infty$), Eqs.(15,16) become 
\be
\Sigma(\omega)=\int_{\omega^{\prime}}
{\Lambda(\omega,\omega^{\prime})\over \Pi(\omega+\omega^{\prime})} 
{\Sigma(\omega^{\prime})\over \Sigma^2(\omega^{\prime})+|\Phi(\omega^{\prime})|^2}
\label{eq:sc-sigma}
\ee
and
\be
\Phi(\omega)=\int_{\omega^{\prime}}
{\Lambda(\omega,\omega^{\prime})\over \Pi(\omega+\omega^{\prime})} 
{\Phi(\omega^{\prime})\over \Sigma^2(\omega^{\prime})+|\Phi(\omega^{\prime})|^2}
\label{eq:sc-phi}
\ee
while the polarization simplifies down to 
\bea
\Pi(\omega)=\int_{\omega^{\prime}}
\Lambda(\omega+\omega^{\prime},\omega^{\prime})~~~~~~~~~~~~\\
{\Sigma(\omega^{\prime})\Sigma(\omega+\omega^{\prime})-\Phi^{*}(\omega^{\prime})\Phi(\omega+\omega^{\prime})
\over (\Sigma^2(\omega+\omega^{\prime})+|\Phi(\omega+\omega^{\prime})|^2)
(\Sigma^2(\omega^{\prime})+|\Phi(\omega^{\prime})|^2)}
\nonumber
\label{eq:sc-pi}
\eea 
Notably, Eqs.(18) and (19) can be combined into a single one for a complex-valued 
function $\Psi=\Sigma+i\Phi$ which obeys the intrinsically non-linear equation
\be
\Psi(\omega)=\int_{\omega^{\prime}}
{\Lambda(\omega+\omega^{\prime},\omega^{\prime}) \over \Pi(\omega+\omega^{\prime})} 
{1\over \Psi(\omega^{\prime})}
\label{eq:psi-unified}
\ee
thus making the symmetry between the particle-hole (signified by $\Sigma$) and particle-particle (signified by $\Phi$) channels manifest. 

Considered at its face value, Eq.(21) demonstrates that in the ultra-local infinite-coupling limit 
the system develops an effective symmetry (and, therefore, fails to discriminate) 
between a non-quasiparticle NFL behavior and superconducting pairing,
both contributing equally to the total zero-frequency gap $\Delta={\sqrt {\Sigma^2(0)+\Phi^2(0)}}$.  
While such a non-separability would be quite natural to expect for real-valued ($\psi={\bar \psi}$) Majorana fermions, the analogous behavior of their Dirac counterparts may require a further insight.  
\\

\noindent
{\it Normal state self-energy}
\\

\noindent 
In the normal state, by putting $\Phi=0$ in Eq.(21) and applying the Ward identity
\be
(\omega-\omega^{\prime})\Lambda(\omega,\omega^{\prime})=
G^{-1}(\omega)-G^{-1}(\omega^{\prime})
\label{eq:ward}
\ee
one finds that the frequency integral in (21) vanishes identically if the propagator $G(\omega)$ 
has a pure pole structure. In this situation, the fermion polarization is absent and the 
ultra-strong limit becomes unattainable, so the system remains stuck
in the weak-coupling regime where $d(\omega)\approx D_0(\omega)$.

However, in the general case of a branch-cut singularity the polarization (20) 
still remains (potentially) non-trivial while Eq.(21) assumes the universal form  
\be
1=\int_{\omega^{\prime}}
{
G(\omega^{\prime})-G(\omega)\over 
{\int_{\omega^{\prime\prime}}
(G(\omega-\omega^{\prime}+\omega^{\prime\prime})
-G(\omega^{\prime\prime})) 
}
}
\label{eq:universal-normal}
\ee
which hints at a (nearly) frequency-independent 
(yet, strongly cut-off-dependent) self-energy $\Sigma(\omega)=const$, albeit not determining its absolute value.  
In order to obtain a non-trivial solution with the properties
$\Sigma(\omega)/\omega\to\infty$ at $\omega\to 0$ 
and $\Sigma(\omega)/\omega\to 0$ at $\omega\to \infty$ 
one has to restore the bare term $\omega$ in Eq.(4).
A systematic analysis of this equation will be presented elsewhere \cite{khveshchenko_unpublished}.

Alternatively, one can gain some preliminary insight into the self-consistent solutions of (21) 
characterized by a power-law behavior of the polarization $\Pi(\omega)\sim |\omega|^{\gamma}$. 
Utilizing a like ansatz 
$\Sigma\sim\omega^{\zeta}$  one finds $\zeta=(1-\gamma)/2$ which value should be contrasted against its counterpart in the case of fermions with a non-trivial dispersion where the exponent $\zeta$ would be twice larger,
thus suggesting a generally stronger deviation towards NFL in the flat-band limit where kinetic energy is strongly quenched.  

Alternatively, for $\Phi=0$ Eq.(18) can be converted into a (non-linear) 
differential equation for the normal-state self-energy $\Sigma(\omega)$. 
Similar procedures have been repeatedly exercised 
in a variety of problems where such integral equations arise for,    
both, Lorentz-invariant and manifestly non-invariant (i.e., purely momentum- 
or energy-) dependent functions, 
including chiral symmetry breaking in QED \cite{appelquist_spontaneous_1986}, 
CDW ordering in graphene \cite{khveshchenko_ghost_2001,gorbar_magnetic_2002,leal_excitonic_2004,khveshchenko_excitonic_2006,khveshchenko_massive_2009}, as well as 
superconducting instabilities in the NFL metals  \cite{abanov_interplay_2020,wu_interplay_2020,wu_interplay_2020-1,wu_interplay_2021,wu_interplay_2021-1,zhang_interplay_2021,yu_topologically_2026}.

Following the procedure described in the above references one then obtains 
\be
{d\over d\omega}(\omega^{1+\gamma}{d\over d\omega}\Sigma)+{g\gamma\over \Sigma}=0
\label{eq:sigma-ode}
\ee
For $\gamma=-1$ one can easily find the first integral 
\be
I={1\over 2}({d\Sigma\over d\omega})^2-g\ln({\Omega\over \Sigma})
\label{eq:first-integral}
\ee
where the Debye frequency $\Omega$ is used as the ultraviolet (UV) cut-off.
With the use of (25) one then obtains a formal solution given by the quadrature  
\be
\omega={\sqrt 2}\int^{\Sigma_0}_0{d\Sigma\over {\sqrt {I+g\ln(\Sigma)}}}
\label{eq:quadrature}
\ee
For $|\gamma+1|<<1$ the behavior described by Eq.(24) resembles that of a weakly dissipative (generically, non-Ohmic) Hamiltonian system.
Its solutions can then be found by applying classical perturbation theory
to the non-dissipative solution $\Sigma_0(\omega)$ given by Eq.(26) \cite{khveshchenko_unpublished}.

However, it was recently argued that a breakdown of the ME approximation in the normal state may occur even 
at moderate couplings due to the intrinsic (bi-)polaronic effects \cite{migdal_interaction_1958,eliashberg_interactions_1960,eliashberg_temperature_1963,alexandrov_breakdown_2001,marsiglio_eliashberg_2020,bauer_quantitative_2011,chowdhury_unreasonable_2020,yang_non-perturbative_2023,pan_nonperturbative_2021,huang_theoretical_2023,huang_superconductivity_2025,schrodi_full-bandwidth_2020,damia_thermal_2021,wang_non-fermi_2017,wang_superconductivity_2018,yuzbashyan_superconductivity_2022,yuzbashyan_breakdown_2022,yuzbashyan_migdal-eliashberg_2022,yuzbashyan_instability_2025,kiessling_bounds_2025,kiessling_bounds_2025-1,kiessling_bounds_2025-2,
semenok_fundamental_2025} which has been ignored in the above analysis.

For the sake of completeness, it is worth presenting the self-energy in the polaron regime 
where it takes a closed form of the continued fraction \cite{paci_polaronic_2005,fratini_optical_2006,heidarian_inhomogeneous_2003,wolfle_self-consistent_2010,miskic_polaron_2025,ragni_polarons_2025,chubukov_breakdown_2026,prokofev_limits_2026,
chubukov_eliashberg_2020,cai_superconductivity_2025,kabanov_adiabatic_2020}
\be
\Sigma(\omega)={g^{2}G_0(\omega-\Omega)\over {1-{2g^2G_0(\omega-\Omega)G_0(\omega-2\Omega)
\over {1-3g^2G_0(\omega-2\Omega)G_0(\omega-3\Omega)}}}\dots}
\label{eq:eikonal-cf}
\ee
which result was obtained by virtue of the so-called eikonal approximation  \cite{paci_polaronic_2005,fratini_optical_2006,heidarian_inhomogeneous_2003,wolfle_self-consistent_2010,miskic_polaron_2025,ragni_polarons_2025,chubukov_breakdown_2026,prokofev_limits_2026,
chubukov_eliashberg_2020,cai_superconductivity_2025,kabanov_adiabatic_2020}. It 
specifically targets the family of crossed diagrams and, therefore, its results 
appear to be expressly different from those obtained in the framework of the non-crossing ME theory.

Thus-found self-energy further sums up into the series 
\be
G(\omega)=e^{-(g/\Omega)^2}\sum^{\infty}_{n=0}
({g\over \Omega})^{2n}{1\over n!}{1\over \omega+g^2/\Omega-n\Omega+i\delta}
\label{polaron-g}
\ee
The spectrum read off of the poles 
at $\omega=-g^2/\Omega+n\Omega$ features a shift of the chemical potential (for $n=0$)
and a tower of equidistant levels for $n\in N$, the corresponding  
DOS being given by the infinite serious of the evenly spaced $\delta$-functions.
While properly accounting for the renormalizing effect of the surrounding phonon cloud on the high-energy  
electron spectra, Eq.(28) lacks the subtle NFL features and concomitant pairing correlations in the immediate vicinity of
the Fermi surface. \\

\noindent
{\it Linearized gap equation}
\\

\noindent
The onset of BCS pairing in the NFL normal state can also be studied by linearizing 
Eq.(19) for the gap function, thus reducing it to some (linear) eigen-function integral equation.
Assuming that the fermion propagator and self-energy conform to the algebraic ansatz
\be
G(\tau)=A{sgn\tau\over \tau^{2\Delta_{\psi}}},~~~~~~~~~~\Sigma(\tau)={1\over A}
{sgn\tau\over \tau^{2-2\Delta_{\psi}}}
\label{eq:conformal-ansatz}
\ee
one can match the parameter $\gamma$ with the fermion dimension $\Delta_{\psi}$ as $\gamma=4\Delta_{\psi}-1$.  
After linearization Eq.(19) takes the form    
\be
\Phi(\omega)=g\int_{\omega^{\prime}}
{\Phi(\omega^{\prime})\over {|\omega^{\prime}|^{1-\gamma}|\omega-\omega^{\prime}|^{\gamma}}}
\label{gap-linear}
\ee
By analogy with the above analysis of the normal-state self-energy, the common analytical approach 
to solving the eigenvalue integral equation (30) follows the route of turning it into a differential one
\be
{d^2\over d\omega^2}\Phi+{\gamma+1\over \omega}{d\over d\omega}\Phi+{g\gamma\over \omega^2}\Phi=0
\label{gap-ode}
\ee 
Seeking 
algebraic solutions in the form $\Phi(\omega)\sim\omega^{\eta-1}$, one then 
obtains a quadratic equation for the exponent, with the roots  
\be
\eta_{\pm}=1-{1\over 2}\gamma\pm i{\sqrt {g\gamma-{1\over 4}\gamma^2}}
\label{eta-roots}
\ee
exhibiting a critical coupling $g_c=\gamma/4$ above which (32) becomes complex-valued. 

At small frequencies and/or temperatures one expects 
that solutions of the original integral equation (30) to approach a constant. 
Further imposing a boundary condition forcing the gap function
$\Phi_0(\omega)$ to vanish at the UV 
cut-off $\Omega$ and level off at low frequencies $\omega\lesssim\Delta$ 
results in the interpolating formula \cite{khveshchenko_ghost_2001,gorbar_magnetic_2002,leal_excitonic_2004,khveshchenko_excitonic_2006,khveshchenko_massive_2009}  
\be
\Phi_0(\omega)={\Delta^{1+\gamma/2}\over ({\omega+\Delta})^{\gamma/2}}
\sin({\sqrt {g-g_c}}\ln{\Omega+\Delta\over \omega+\Delta})
\label{eq:gap-interp}
\ee 
The amplitude of the gap function is then controlled by the low-frequency (infrared or IR) asymptotic value   
\be
\Delta=\Omega\exp(-{\pi\over 2{\sqrt {g-g_c}}})
\label{eq:gap-amplitude}
\ee
which demonstrates a characteristic 'conformal' critical 
behavior as a function of the coupling $g$ 
(cf. the Kosterlits-Thouless formula at finite temperatures). 
Unlike in the standard BCS scenario, Eq.(34) suggests that pairing emerges  
out of the 'ultra-local' NFL ground state only if the coupling strength exceeds a certain threshold. 

Notably, on top of the monotonically decreasing (hence, sign-preserving) minimal
solution, $d\Phi_0(\omega)/d\omega<0$, the differential 
Eq.(31) possesses an infinite set of non-monotonic  
ones with $n$ additional 
nodes within the interval $0<\omega<\Omega$. 
The amplitudes of such oscillating functions $\Phi_n(0)\sim\Delta\exp(-\pi n/{\sqrt {g-g_c}})$ 
are progressively more and more strongly suppressed as compared to the monotonic one. In Refs.\cite{abanov_interplay_2020,wu_interplay_2020,wu_interplay_2020-1,wu_interplay_2021,wu_interplay_2021-1,zhang_interplay_2021,yu_topologically_2026}
they were analyzed in all the excruciating details.
 
It has been, however, reported that the condensation energy (6) 
attains its lowest value on the $n=0$ solution for all $\gamma<2$ \cite{tsoncheva_condensation_2005,marsiglio_electron_2001,mirabi_thermodynamics_2020,
zhang_superconductivity_2022,zhang_free_2023-1,abanov_interplay_2020,wu_interplay_2020,
wu_interplay_2020-1,wu_interplay_2021,wu_interplay_2021-1,zhang_interplay_2021,yu_topologically_2026}. 
Therefore, it is highly likely that, in contrast to the monotonic ($n=0$)  
solution (33), all the non-monotonic ones with $n>0$ turn out to be spurious and not readily realizable. 
\cite{paci_polaronic_2005,fratini_optical_2006,heidarian_inhomogeneous_2003,wolfle_self-consistent_2010,miskic_polaron_2025,ragni_polarons_2025,chubukov_breakdown_2026,prokofev_limits_2026,
chubukov_eliashberg_2020,cai_superconductivity_2025,kabanov_adiabatic_2020,yuzbashyan_superconductivity_2022,
yuzbashyan_breakdown_2022,yuzbashyan_migdal-eliashberg_2022,yuzbashyan_instability_2025,kiessling_bounds_2025,kiessling_bounds_2025-1,
kiessling_bounds_2025-2,semenok_fundamental_2025}.  
\\ 
 
\noindent
{\it SYK model}
\\

\noindent
The Hamiltonian of the original real/complex $SYK_q$ model is given by the sum over equally-weighted products of an even number $q$ of the Majorana \cite{Kitaev2015,kitaev_notes_2018,kitaev_soft_2018,kitaev_statistical_2019,sachdev_holographic_2010,sachdev_bekenstein-hawking_2015,sachdev_strange_2023,maldacena_bound_2016,maldacena_remarks_2016,maldacena_conformal_2016,stanford_fermionic_2017,polchinski_spectrum_2016,gross_generalization_2017,gross_all_2017,sarosi_ads_2_2018,bagrets_sachdev-ye-kitaev_2016,bagrets_power-law_2017,mertens_solving_2017,mertens_schwarzian_2018,yang_quantum_2019,jha_theory_2026,khveshchenko_thickening_2018,khveshchenko_seeking_2018,khveshchenko_true_2019,khveshchenko_one_2020,khveshchenko_gloria_2022}
or Dirac \cite{tikhanovskaya_excitation_2021,gu_notes_2020,pethybridge_notes_2024,zhang_gauging_2025,gubankova_scaling_2026,solis_single-particle_2025,akyuz_four-point_2026} fermions 
\be
H_{syk}=i^{q/2}J\sum_{1\leq<i_1<\dots <i_{q}\leq N}
J_{i_1,\dots,i_{q}}\psi_{i_1}\dots,\psi_{i_{q}}
\label{eq:syk-hamiltonian}
\ee
where the couplings $J$ are drawn from the Gaussian random ensemble with the variance
\be
<J_{i_1,\dots,\i_q}J_{i^{\prime}_1,\dots,\i^{\prime}_q}>=
{J^2(q-1)!\over N^{q-1}}\prod_{k=1}^N\delta_{i_k,i^{\prime}_k}
\label{eq:syk-variance}
\ee
It has been extensively discussed in the context of the properties of the normal NFL states of 
$N>>1$ dispersion-less fermions.

At large $N$, summation of the dominant melonic graphs yields a self-energy \cite{Kitaev2015,kitaev_notes_2018,kitaev_soft_2018,kitaev_statistical_2019,sachdev_holographic_2010,sachdev_bekenstein-hawking_2015,sachdev_strange_2023,maldacena_bound_2016,maldacena_remarks_2016,maldacena_conformal_2016,stanford_fermionic_2017,polchinski_spectrum_2016,gross_generalization_2017,gross_all_2017,sarosi_ads_2_2018,bagrets_sachdev-ye-kitaev_2016,bagrets_power-law_2017,mertens_solving_2017,mertens_schwarzian_2018,yang_quantum_2019,jha_theory_2026,khveshchenko_thickening_2018,khveshchenko_seeking_2018,khveshchenko_true_2019,khveshchenko_one_2020,khveshchenko_gloria_2022,klebanov_uncolored_2017,peng_vector_2017,fraser-taliente_melonic_2024}
\be
\Sigma(\tau)=qJ^2G^{q-1}(\tau)
\label{eq:melonic-sigma}
\ee
which, in turn, gives rise to the Luttinger-Ward (LW) $'G-\Sigma'$ potential
\bea
F_{syk}[G,\Sigma]=-N\ln Pf(\partial_{\tau}-\Sigma)\\
+\int_{\tau} ({J^2}G^q(\tau)-N\Sigma(\tau)G(\tau)) \nonumber
\label{eq:lw-syk}
\eea
By direct inspection one finds that the SD equation with the self-energy
(37) can be readily solved by the algebraic ('conformal') ansatz (29) 
provided that the fermion dimension takes the value $\Delta_{\psi}=1/q$ \cite{Kitaev2015,kitaev_notes_2018,kitaev_soft_2018,kitaev_statistical_2019,sachdev_holographic_2010,sachdev_bekenstein-hawking_2015,sachdev_strange_2023,maldacena_bound_2016,maldacena_remarks_2016,maldacena_conformal_2016,stanford_fermionic_2017,polchinski_spectrum_2016,gross_generalization_2017,gross_all_2017,sarosi_ads_2_2018,bagrets_sachdev-ye-kitaev_2016,bagrets_power-law_2017,mertens_solving_2017,mertens_schwarzian_2018,yang_quantum_2019,jha_theory_2026,khveshchenko_thickening_2018,khveshchenko_seeking_2018,khveshchenko_true_2019,khveshchenko_one_2020,khveshchenko_gloria_2022}.

In the dispersion-less case of the $1+0$-dimensional 'ostensible' FL the formal fermion 
dimension is $\Delta_{\psi}=0$ while for an extended 
Fermi surface in $1+d$-dimensions it takes on the universal value $\Delta_{\psi}=1/2$ for any $d>0$. 
Notably, $\Delta_{\psi}$ differs from either value for all $q\geq 4$ which guarantees 
that the $q$-fermion coupling term in the SYK action remains exactly marginal and, therefore, dominant over the kinetic one, $\int_{\tau} \psi\partial_{\tau}\psi$, of the dimension $2/q>0$. 

On a side note, in the large-$q$ limit the normal-state fermion function  
can be found by utilizing the exponential representation \cite{Kitaev2015,kitaev_notes_2018,kitaev_soft_2018,kitaev_statistical_2019,sachdev_holographic_2010,sachdev_bekenstein-hawking_2015,sachdev_strange_2023,maldacena_bound_2016,maldacena_remarks_2016,maldacena_conformal_2016,stanford_fermionic_2017,polchinski_spectrum_2016,gross_generalization_2017,gross_all_2017,sarosi_ads_2_2018,bagrets_sachdev-ye-kitaev_2016,bagrets_power-law_2017,mertens_solving_2017,mertens_schwarzian_2018,yang_quantum_2019,jha_theory_2026,khveshchenko_thickening_2018,khveshchenko_seeking_2018,khveshchenko_true_2019,khveshchenko_one_2020,khveshchenko_gloria_2022}
\bea
G_{\psi}(\tau)={sgn\tau\over 2} e^{g_{\psi}/q}={sgn\tau\over 2}(1+g_{\psi}/q+...),\nonumber\\
\Sigma_{\psi}(\omega)={\omega^2\over 2q}[sgn\tau g_{\psi}]_{\omega}+\dots
\label{eq:large-q-ansatz}
\eea
where the function $g_{\psi}(\tau_1, \tau_2)$ satisfies the Liouville equation
\be
\partial_{\tau}^2g_{\psi}-\partial_T^2g_{\psi}-2J^2sgn\tau\exp(g_{\psi})=0
\label{eq:liouville}
\ee
where $T=(\tau_1+\tau_2)/2$ and $\tau=\tau_1-\tau_2$. Switching to the variables $T$ and $\tau$ exposes the intrinsic  $1+1$-dimensional kinematic structure of the bi-local quantities in question and makes it tempting to explore 
the possibility of a concomitant (pseudo-)holographic connection (see below).
\\

\noindent
{\it YSYK model}
\\

\noindent
Non-Gaussian (quartic, etc.) fermion couplings can be alternatively described by introducing an auxiliary phonon-like boson field which mediates interactions
between the Majorana/Dirac fermions, thus resulting 
in the real/complex 
YSYK model, respectively.

Furthermore, the latter allows for an immediate generalization to an arbitrary type of the local coupling  
between $q$ fermions and $p$ bosons (of $N$ and $M$ flavors, respectively) 
\bea
L_{ysyk}=\sum_{i=1}^N\psi_i\partial_{\tau}\psi_i+{1\over 2}\sum_{\alpha=1}^M
(\Omega^2\phi^2_{\alpha}+(\partial_{\tau}\phi_{\alpha})^2)~~~~~~~~~~~~\\
+i\sum_{1\leq\alpha_1\leq\dots\leq\alpha_p\leq M}
\sum_{1\leq i_1<\dots <i_q\leq N}
g_{\alpha_{1}\dots{\alpha}_{p};i_1\dots i_{q}}
\prod^{p}_{k=1}\phi_{\alpha_k}
\prod^{q}_{j=1}\psi_{j_l}\nonumber
\label{eq:ysyk-lagrangian}
\eea
with the random coupling amplitudes $g_{\alpha_{1}\dots{\alpha}_{p}i_1\dots i_{q}}$.
 
In the normal state, the fermion self-energy and boson polarization 
\be
\Sigma(\tau)=g^2qG^{q-1}(\tau)D^p(\tau),~~\Pi(\tau)=g^2pD^{p-1}(\tau)G^q(\tau)
\label{eq:ysyk-melonic}
\ee
determine the generalized LW (or $'G/D-\Sigma/\Pi'$) functional
\bea
F_{ysyk}[G,\Sigma,D,\Pi]=
-N\ln Pf(\partial_{\tau}-\Sigma)+{M\over 2}\ln(1-\Omega\Pi)\nonumber\\
+\int_{\tau}(gD^p(\tau)G^{q}(\tau)-NG(\tau)\Sigma(\tau)+{M\over 2}D(\tau)\Pi(\tau))~~~~~~~~~
\label{eq:lw-ysyk}
\eea 
Akin to the algebraic SYK ansatz (29), the normal state of the YSYK theory (41) 
has been extensively studied in the conformal regime where, both, the fermion and boson 
propagators are assumed to behave algebraically in the time domain \cite{chubukov_superconductivity_2005,bi_instability_2017,classen_superconductivity_2021,wang_phase_2021,hauck_eliashberg_2020,wang_solvable_2020,esterlis_cooper_2019,chowdhury_intrinsic_2020,pan_yukawa-syk_2021,aldape_solvable_2022,choi_pairing_2022,chowdhury_sachdev-ye-kitaev_2022,wang_quantum_2020,bashan_tunable_2024,li_strange_2024,davis_quantum_2023,valentinis_correlation_2023,valentinis_bcs_2023,lantagne-hurtubise_superconducting_2021,zhang_density_2023,elezaby_superconductivity_2026,gnezdilov_upper_2025,sutradhar_singlet_2024,bonetti_fractionalized_2026,kleger_universal_2026}
\be
G(\tau)=A{sgn\tau\over \tau^{2\Delta_{\psi}}}, ~~~~~~~~~D(\tau)=B{1\over \tau^{2\Delta_{\phi}}}
\label{eq:ysyk-ansatz}
\ee 
Plugging Eqs.(44) into either of Eqs.(42)  one finds the consistency condition
\be 
q\Delta_{\psi}+p\Delta_{\phi}=1
\label{eq:dim-sum}
\ee
The minimal $YSYK_q$ model describing the basic two-fermion interaction mediated by a single boson exchange is recovered for $p=1$ and $q=2$. With the use of Eq.(45) one arrives at the single-parameter ansatz 
\be
G(\tau)=A{sgn\tau\over \tau^{2\Delta_{\psi}}}, ~~~~~~~~~D(\tau)=B{1\over \tau^{2-4\Delta_{\psi}}}
\label{eq:ansatz-minimal}
\ee 
which equations are consistent with taking $g$ 
to the strong coupling regime (as in Eqs.(18-20)  where the renormalized bosonic propagator behaves as 
$D(\omega)\sim 1/\Pi(\omega)$.

In the opposite weak-coupling limit, the instantaneous bare boson propagator 
$D_0(\omega)=1/\Omega$  
gets dressed by the polarization $\Pi(\omega)$ as  
\be
D(\omega)={D_{0}\over {1+gD_{0}\Pi(\omega)}}=
{1\over \Omega}-{g\over \Omega^{2}}\Pi(\omega)+\dots
\label{eq:boson-dressed}
\ee
where the lowest order term does not contribute to Eqs.(42) due to the UV 
regularization imposed by the condition $G(0)=0$. 

Interestingly, functional integration over the boson field  restores the Hamiltonian (35) of the original $SYK_{2q}$ model (note the doubling of the factor $q$). In particular, the solution (29) still holds, as per the weak-coupling relation $D(\tau)\sim G^{q-3}(\tau)G(-\tau)$, consistent with Eq.(46).

Utilizing the ansatz Eq.(46), computing the integrals in either of Eqs.(42) (see Appendix),  
and equating the coefficients on both sides  under the assumption of non-zero prefactors $A$ and $B$ 
(which remain undetermined) produces the relation
\be
{N\over M{q}}(1-2\Delta_{\psi})\tan(\pi\Delta_{\psi})=(1-2\Delta_{\phi})\cot(\pi\Delta_{\phi})
\label{eq:dim-relation}
\ee
Notably, in some of the original works on YSYK \cite{chubukov_superconductivity_2005,bi_instability_2017,classen_superconductivity_2021,wang_phase_2021,hauck_eliashberg_2020,wang_solvable_2020,esterlis_cooper_2019,chowdhury_intrinsic_2020,pan_yukawa-syk_2021,aldape_solvable_2022,choi_pairing_2022,chowdhury_sachdev-ye-kitaev_2022,wang_quantum_2020,bashan_tunable_2024,li_strange_2024,davis_quantum_2023,valentinis_correlation_2023,valentinis_bcs_2023,lantagne-hurtubise_superconducting_2021,zhang_density_2023,elezaby_superconductivity_2026,gnezdilov_upper_2025,sutradhar_singlet_2024,bonetti_fractionalized_2026,kleger_universal_2026} 
one finds the (incorrect) variant of Eq.(48) where 
the factor $\cot\pi\Delta_{\phi}$ is replaced by $\tan\pi\Delta_{\phi}$. 
Another erroneous form of the formula (48) can be found in the recent review \cite{abchub} 
that appeared after the release of the original version of the current manuscript.

It can be readily seen that for $N/M=1$ (which ratio corresponds to the supersymmetric (SUSY) 
case, see below) Eq.(48) allows for the analytic solution with rational dimensions  
\be 
\Delta_{\psi}=1/6,~~~~~\Delta_{\phi}=2/3
\label{eq:dim-susy-min}
\ee
For comparison, at $M/N>>1$ Eq.(48) shows that the fermion dimension approaches 
the standard $SYK_4$ result $\Delta_{\psi}=1/4$,  
while in the opposite limit ($M/N<<1$) it attains the FL value $\Delta_{\psi}=1/2$.

At large $q$ the SD equations in the YSYK model can also be solved with the substitution (39) 
and an analogous formula for the boson propagator 
$
D(\tau)=-\delta(\tau)+g_{\phi}/2q+...
$, thereby further elaborating on the aforementioned connection to the $1+1$-dimensional
Liouville theory. This link will be pursued elsewhere \cite{khveshchenko_unpublished}.
\\

\noindent
{\it SUSY (Y)SYK models}
\\

\noindent
By construction, the Hamiltonians of the SUSY generalizations of the SYK model
are given by anticommutators of a number ${\cal N}$ of supercharges \cite{fu_supersymmetric_2017,peng_soft_2021,biggs_supersymmetric_2024,heydeman_phases_2023,gates_1d_2021,peng_correlators_2017,marcus_new_2019,heydeman_supersymmetry_2024,murugan_more_2017,benini_mathcaln2_2025,berkooz_double_2020}.
The minimal (${\cal N}=1$) SUSY SYK features a single supercharge 
\be
Q=i^{({\hat q}-1)/2}\sum_{1\leq i_1<\dots <i_{\hat q}\leq N} C_{i_1,\dots, i_{\hat q}}
\psi_{i_1}\dots\psi_{i_{\hat q}}
\label{eq:supercharge}
\ee
given by the product of an odd number ${\hat q}\geq 3$ of Majorana fermions, 
their Hamiltonian being $H_{{\cal N}=1}=Q^2$. 
Contrary to the original SYK model where the Hamiltonian parameter $J$ itself is Gaussian random, 
the ensemble averaging now involves squares of the random amplitudes with the variance 
\be
<C_{i_1,\dots, i_{\hat q}}C^{*}_{i^{\prime}_1,\dots, i^{\prime}_{\hat q}}> 
=
{J^{\hat q}\over {\hat q}{!}{\hat q}N^{{\hat q}-1}}\prod_{k=1}^N\delta_{i_ki^{\prime}_k}
\label{eq:susy-variance}
\ee
Upon averaging one arrives at the $SYK_q$ model with  
\be
q=2({\hat q}-1)
\label{eq:q-map}
\ee
where the reduction in the number of the locally interacting fermions by two units stems from anti-commutation. 
Likewise, the ${\cal N}=2$ variant with the Hamiltonian
$H_{{\cal N}=2}={\{ }Q,Q^{\dagger}{\} }$ is constructed out of the 
Dirac - rather than Majorana - fermions. 

At the Lagrangian level the ${\cal N}=2$ theory allows for two equivalent representations.
One is in terms of the chiral fermionic superfield (here $\theta$ is a super-coordinate) 
\be
\Psi(\tau,\theta,{\bar \theta})=\psi(\tau+{\bar \theta}\theta)+\theta\phi
\label{eq:superfield-fermionic}
\ee
subject to the constraint $D_{\bar \theta}\Psi=0$, where $D_{\bar {\theta}}=\partial_{{\bar \theta}}+\theta\partial_{\tau}$,
and its conjugate ${\bar \Psi}(\tau,{\theta},{\bar \theta})={\bar \psi}+{\bar \theta}\phi^{*}$ obeying $D_{\theta}{\bar \Psi}=0$ where $D_{{\theta}}=\partial_{{\theta}}+{\bar \theta}\partial_{\tau}$ \cite{fu_supersymmetric_2017,peng_soft_2021,biggs_supersymmetric_2024,heydeman_phases_2023,gates_1d_2021,peng_correlators_2017,marcus_new_2019,heydeman_supersymmetry_2024,murugan_more_2017,benini_mathcaln2_2025,berkooz_double_2020}.

It is then described by the SUSY Lagrangian 
\bea
L_{{\cal N}=2}=\int d\theta d{\bar \theta}{1\over 2}{\bar \Psi}\Psi\nonumber\\ 
+ {i^{({\hat q}-1)/2}}
\int d\theta C_{i_1\dots i_{\hat q}}\Psi_{i_1}\dots\Psi_{i_{\hat q}}+h.c.
\label{eq:lag-n2-super}
\eea
which can be equivalently formulated with the use of a minimally coupled bosonic mediator  
($p=1$ in Eq.(41)), so that the counterpart of Eq.(45) takes the form 
\be 
({\hat q}-1)\Delta_{\psi}+\Delta_{\phi}=1
\label{eq:dim-sum-susy}
\ee  
In terms of the component fields the Lagrangian (53) reads 
\bea
L_{{\cal N}=2}=\sum_{i=1}^N({\bar \psi}_{i}\partial_{\tau}\psi_{i}
-{1\over 2}\Omega^2{\bar \phi}_i\phi_{i})+~~~~~~~~~~~\nonumber\\
i^{({\hat q}-1)/2} 
\sum_{j;1\leq j_1\dots <j_{{\hat q}-1}\leq N}
C_{jj_1\dots j_{{\hat q}-1}}
{\bar \phi}_j{\psi}_{j_1}\dots\psi_{j_{{\hat q}-1}}~~\nonumber\\
\label{eq:lag-n2-comp}
\eea
In particular, choosing the minimal value ${\hat q}=3$ and adding 
a squared time derivative of the boson field (which is an irrelevant perturbation in the IR) reproduces the special ($M/N=1$) case of the YSYK Lagrangian (41) for the basic 
two-fermion interaction facilitated by a single boson exchange.
Conversely, for an equal number of the fermion and boson species 
the YSYK model (41) of dispersion-less electrons interacting via optical phonons 
develops ${\cal N}=2$ SUSY in the asymptotic IR regime.  
  
Alternatively, one can construct the ${\cal N}=2$ theory   
in terms of the bosonic supermultiplet \cite{fu_supersymmetric_2017,peng_soft_2021,biggs_supersymmetric_2024,heydeman_phases_2023,gates_1d_2021,peng_correlators_2017,marcus_new_2019,heydeman_supersymmetry_2024,murugan_more_2017,benini_mathcaln2_2025,berkooz_double_2020}
\be
\Psi_i(\tau,\theta_{\alpha})=\phi_i+i{\theta}_{\alpha}\psi_{i\alpha}+
iF_i\epsilon_{\alpha\beta}{\theta}_{\alpha}\theta_{\beta}
\label{eq:superfield-bosonic}
\ee
(hereafter,  a summation over pairs of the two-valued Greek indices is assumed).
This theory is governed by the Lagrangian
\be
L^{\prime}_{{\cal N}=2}=\int d\theta d{\bar \theta}
({i\over 4}\epsilon_{\alpha\beta}D_{\alpha}{\Psi}_{i}D_{\beta}\Psi_{i} + 
C_{i_1\dots i_q}\Psi_{i_1}\dots\Psi_{i_{\hat q}})
\label{eq:lag-n2-boson}
\ee
where $D_{\alpha}=\partial_{\theta_{\alpha}}+{\bar \theta}_{\alpha}\partial_{\tau}$.

After having been cast in terms of the component fields the minimal (${\hat q}=3$) Lagrangian (58) takes the form  
\bea
L^{\prime}_{{\cal N}=2}=\sum_{i=1}^N{1\over 2}{\psi}_{i\alpha}\partial_{\tau}\psi_{i\alpha}+
{1\over 2}\sum(\Omega^2{F}^2_i+(\partial_{\tau}\phi_i)^2)+\nonumber\\
3i\sum_{k;1\leq i< j\leq N}C_{ijk}
(\phi_{i}\phi_{j}F_k+\epsilon_{\alpha\beta}\psi^{\alpha}_i\psi^{\beta}_j\phi_k)~~~~~~~~~~
\label{eq:lag-n2-fterm}
\eea
An attempt to integrate out the bosonic field $F_i$ gives rise to additional 
anharmonic (quartic) phonon terms, their 
amplitudes matching the square of the electron-phonon coupling. 

Moreover, with a separate  two-valued 
'spin' degree of freedom, the theory can be further elevated to the ${\cal N}=4$ SUSY
in terms of the chiral bosonic superfield 
$
{\Psi}_{i\alpha}(\tau,\theta_{\alpha},{\bar \theta}_{\alpha})
$
and its conjugate.
Unlike in the case of ${\cal N}=1$ or $2$, though,
 the ${\cal N}=4$ theory 
requires the additional bosons $F_i^{\alpha}$ 
to be dynamical
fields, rather than mere auxiliary variables (Lagrange multipliers) \cite{fu_supersymmetric_2017,peng_soft_2021,biggs_supersymmetric_2024,heydeman_phases_2023,gates_1d_2021,peng_correlators_2017,marcus_new_2019,heydeman_supersymmetry_2024,murugan_more_2017,benini_mathcaln2_2025,berkooz_double_2020}.

In the YSYK model (41) with a generic (rational) value of the ratio $M/N$, 
the boson and fermion sectors can have rather different properties 
(thermodynamics, kinetic coefficients, etc.) By contrast, in the SUSY YSYK model 
the fermion and boson dynamics are tied together. 

In particular, the unbroken ${\cal N}=2$ SUSY imposes a stringent constraint 
\be
\Delta_{\phi}=\Delta_{\psi}+{1\over 2}
\label{eq:susy-shift}
\ee
thereby implying
\be
D(\tau)=\partial_{\tau}G(\tau),~~~~~~~\Sigma(\tau)=\partial_{\tau}\Pi(\tau)
\label{eq:susy-props}
\ee
Likewise, in the ${\cal N}=4$ 
theory the boson correlators obey the relation   
$
<F(\tau)F(0)>=\partial_{\tau}<\phi\phi>(\tau)
$ \cite{fu_supersymmetric_2017,peng_soft_2021,biggs_supersymmetric_2024,heydeman_phases_2023,gates_1d_2021,peng_correlators_2017,marcus_new_2019,heydeman_supersymmetry_2024,murugan_more_2017,benini_mathcaln2_2025,berkooz_double_2020}.

From Eqs.(55) and (60) and for any ${\hat q}\geq 3$ one then finds the SUSY solution 
\be 
\Delta_{\psi}={1\over {2\hat q}},~~~~~\Delta_{\phi}={1\over 2}+{1\over 2{\hat q}}
\label{eq:dim-susy-general}
\ee
In particular, for ${\hat q}=3$ one obtains the dimensions (49), as suggested by the emerging SUSY of the properly tuned YSYK model. 
Similar to the situation in the original SYK model,
at such dimension values the fermion-boson coupling 
term remains marginal and dominates over the kinetic ones in the IR.

Interestingly, the fermion dimension in the SUSY YSYK falls outside the previously anticipated  
interval $\Delta_{\psi}\in (1/4, 1/2)$ \cite{chubukov_superconductivity_2005,bi_instability_2017,classen_superconductivity_2021,wang_phase_2021,hauck_eliashberg_2020,wang_solvable_2020,esterlis_cooper_2019,chowdhury_intrinsic_2020,pan_yukawa-syk_2021,aldape_solvable_2022,choi_pairing_2022,chowdhury_sachdev-ye-kitaev_2022,wang_quantum_2020,bashan_tunable_2024,li_strange_2024,davis_quantum_2023,valentinis_correlation_2023,valentinis_bcs_2023,lantagne-hurtubise_superconducting_2021,zhang_density_2023,elezaby_superconductivity_2026,gnezdilov_upper_2025,sutradhar_singlet_2024,bonetti_fractionalized_2026,kleger_universal_2026}. In particular, for $q\to\infty$ the fermion dimension becomes 
arbitrarily small, thus further challenging the above expectation.

On the practical side, a small $\Delta_{\psi}$ 
might be beneficial for the superconductivity as it, both, keeps the $dispersion-less$ fermionic quasiparticles from loosing their coherence (hence,  availability for pairing) and also works towards increasing the DOS which, in turn, strengthens the coupling and raises the gap (and/or $T_c$).
For comparison, the solutions of the ${\cal N}=4$ SUSY model features still lower (including negative) dimensions 
$\Delta_{\psi}=-1/3,~~\Delta_{\phi}=2/3,~~\Delta_F=-4/3$ \cite{fu_supersymmetric_2017,peng_soft_2021,biggs_supersymmetric_2024,heydeman_phases_2023,gates_1d_2021,peng_correlators_2017,marcus_new_2019,heydeman_supersymmetry_2024,murugan_more_2017,benini_mathcaln2_2025,berkooz_double_2020} which, however,
might not be immediately applicable to the behavior of any known materials.

It should be pointed out, though, that the previous research on the YSYK model
was primarily focusing on the different (non-SUSY) solutions hallmarked by their irrational dimensions: 
$\Delta_{\psi}=0.42$ (hence $\Delta_{\phi}=0.16$) or $\Delta_{\psi}=0.35$ ($\Delta_{\phi}=0.29$) \cite{chubukov_superconductivity_2005,bi_instability_2017,classen_superconductivity_2021,wang_phase_2021,hauck_eliashberg_2020,wang_solvable_2020,esterlis_cooper_2019,chowdhury_intrinsic_2020,pan_yukawa-syk_2021,aldape_solvable_2022,choi_pairing_2022,chowdhury_sachdev-ye-kitaev_2022,wang_quantum_2020,bashan_tunable_2024,li_strange_2024,davis_quantum_2023,valentinis_correlation_2023,valentinis_bcs_2023,lantagne-hurtubise_superconducting_2021,zhang_density_2023,elezaby_superconductivity_2026,gnezdilov_upper_2025,sutradhar_singlet_2024,bonetti_fractionalized_2026,kleger_universal_2026}. 

The original (non-SUSY) SYK model demonstrates accumulation of the low-energy states but no exact macroscopic ground state degeneracy, despite its zero-temperature (residual) entropy ($S(0)\sim N$ and $S(T)-S(0)\sim T$) computed in the specific (first $N\to\infty$, then $T\to 0$) limit.   
Nor does it have a spectral gap, with DOS vanishing as $\sim E^{1/2}$ \cite{Kitaev2015,kitaev_notes_2018,kitaev_soft_2018,kitaev_statistical_2019,sachdev_holographic_2010,sachdev_bekenstein-hawking_2015,sachdev_strange_2023,maldacena_bound_2016,maldacena_remarks_2016,maldacena_conformal_2016,stanford_fermionic_2017,polchinski_spectrum_2016,gross_generalization_2017,gross_all_2017,sarosi_ads_2_2018,bagrets_sachdev-ye-kitaev_2016,bagrets_power-law_2017,mertens_solving_2017,mertens_schwarzian_2018,yang_quantum_2019,jha_theory_2026,khveshchenko_thickening_2018,khveshchenko_seeking_2018,khveshchenko_true_2019,khveshchenko_one_2020,khveshchenko_gloria_2022}.
In turn, the DOS of the ${\cal N}=1$ theory features a divergence $\sim E^{-1/2}$ at low energies, but still no gap.

By contrast, the unbroken  ${\cal N}=2$ SUSY guarantees a macroscopic    
degeneracy in the ground state - hence the DOS $\sim\delta(E)\exp ({\#}N)$ -    
and a hard excitation gap protected by the conserved Witten index \cite{fu_supersymmetric_2017,peng_soft_2021,biggs_supersymmetric_2024,heydeman_phases_2023,gates_1d_2021,peng_correlators_2017,marcus_new_2019,heydeman_supersymmetry_2024,murugan_more_2017,benini_mathcaln2_2025,berkooz_double_2020}.

In this regard, it is worth mentioning that the minimal ${\cal N}=1$ SUSY
can be spontaneously broken at the non-perturbative level, except in the limit of infinite $N$. On the other hand, the ${\cal N}=2$ SUSY is believed to remain unbroken \cite{fu_supersymmetric_2017,peng_soft_2021,biggs_supersymmetric_2024,heydeman_phases_2023,gates_1d_2021,peng_correlators_2017,marcus_new_2019,heydeman_supersymmetry_2024,murugan_more_2017,benini_mathcaln2_2025,berkooz_double_2020}.

Nonetheless, the ${\cal N}=2$ model is expected to undergo a phase transition  as a function of 
fermion density, akin to that found in the complex SYK model \cite{tikhanovskaya_excitation_2021,gu_notes_2020,pethybridge_notes_2024,zhang_gauging_2025,gubankova_scaling_2026,solis_single-particle_2025,akyuz_four-point_2026}.
In the latter, such transition occurs at a critical density where the conformal solution 
ceases to remain the lowest energy state and the spectrum develops a gap.  
In the complex SYK model, this is the transition between the (high entropy, $S(0)\sim N$) conformal phase 
and the non-conformal (low-entropy, $S(0)\sim 1$) one.
It was also reported that at the same fermion density the ${\cal N}=2$ YSYK model tends to demonstrate
some instability of the  
fermionic excitations themselves, thus suggesting that the two transitions may be closely related \cite{fu_supersymmetric_2017,peng_soft_2021,biggs_supersymmetric_2024,heydeman_phases_2023,gates_1d_2021,peng_correlators_2017,marcus_new_2019,heydeman_supersymmetry_2024,murugan_more_2017,benini_mathcaln2_2025,berkooz_double_2020}.
\\

\noindent
{\it Pairing ladder}
\\

\noindent
Much of the SYK studies centered around computing four-point functions - both, the 
ordinary retarded and out-of-time-order (OTOC) - in the (justified at large $N$) 
ladder approximation. The OTOC functions would often be utilized as markers and 
quantifiers of a chaotic behavior. 
In particular, the thus-discovered maximally chaotic 
nature of the SYK model has been cited as the justification of 
its popular designation as an (approximate) holographic dual of the $1+1$-dimensional
Jackiw-Teitelboim (JT) gravity describing a minimal black hole with the same chaotic properties \cite{Kitaev2015,kitaev_notes_2018,kitaev_soft_2018,kitaev_statistical_2019,sachdev_holographic_2010,sachdev_bekenstein-hawking_2015,sachdev_strange_2023,maldacena_bound_2016,maldacena_remarks_2016,maldacena_conformal_2016,stanford_fermionic_2017,polchinski_spectrum_2016,gross_generalization_2017,gross_all_2017,sarosi_ads_2_2018,bagrets_sachdev-ye-kitaev_2016,bagrets_power-law_2017,mertens_solving_2017,mertens_schwarzian_2018,yang_quantum_2019,jha_theory_2026,khveshchenko_thickening_2018,khveshchenko_seeking_2018,khveshchenko_true_2019,khveshchenko_one_2020,khveshchenko_gloria_2022}. 

Diagrammatically, summing over the ladder diagrams yields solutions to the integral 
eigenvalue equation 
\be
\int K(1,2;3,4)\Psi_{\eta}(3,4)=\lambda(\Delta_{\psi},\eta)\Psi_{\eta}(1,2)
\label{eq:bs-ladder}
\ee 
for the bi-local
operator $\Psi_{\eta}(1,2)$. In the OTOC calculation the latter turns out to be the Casimir operator of the one-dimensional conformal symmetry group $Diff (S^1)/SL(2,R)$ \cite{Kitaev2015,kitaev_notes_2018,kitaev_soft_2018,kitaev_statistical_2019,sachdev_holographic_2010,sachdev_bekenstein-hawking_2015,sachdev_strange_2023,maldacena_bound_2016,maldacena_remarks_2016,maldacena_conformal_2016,stanford_fermionic_2017,polchinski_spectrum_2016,gross_generalization_2017,gross_all_2017,sarosi_ads_2_2018,bagrets_sachdev-ye-kitaev_2016,bagrets_power-law_2017,mertens_solving_2017,mertens_schwarzian_2018,yang_quantum_2019,jha_theory_2026,khveshchenko_thickening_2018,khveshchenko_seeking_2018,khveshchenko_true_2019,khveshchenko_one_2020,khveshchenko_gloria_2022}. 

Likewise, summing over the repeated acts of 
fermion scattering in the Cooper  channel, one obtains the Bethe-Salpeter 
(BS) equation for the pairing order parameter 
represented by the anomalous $<\psi\psi>$ propagator    
with the so-called conformal kernel $K(1,2;3,4)$ 
\bea 
K_{pair}(1,2;3,4)=-{(q-1)(q-2)
\tan{\pi/q}\over 2{\pi}q}\nonumber\\
{sgn(13)sgn(24)\over |13|^{2\Delta_{\psi}}
|24|^{2\Delta_{\psi}}|12|^{2-4\Delta_{\psi}}}
\label{eq:kernel-cooper}
\eea
where $(nm)=\tau_n-\tau_m$.
As in the calculation of the OTOC the possibility of singling out
the ladder diagrams is justified by their dominance for $N>>1$ and is 
consistent with the ME approximation.

Notably, the previous work on the OTOC in the SYK model utilized the kernel 
$ 
K(1,2;3,4)(|12|/|34|)^{2-4\Delta_{\psi}}
$
and its symmetrized version 
$ 
K(1,2;3,4)(|12|/|34|)^{1-2\Delta_{\psi}}
$, 
all of which have the same spectrum as the kernel appearing in the Cooper ladder (64).


The eigenvectors of the kernel (64) can be divided onto the even and odd 
ones under a permutation of their arguments ($\Psi_{\eta}^{+/-}(12)=
\pm\Psi_{\eta}^{+/-}(21)$), so they can be sought out in the algebraic form
\be
\Psi_{\eta}^{+/-}(12)\sim{(1, sgn(12))\over |12|^{2\Delta_{\psi}-\eta}}
\label{eq:eigenvector-ansatz}
\ee
where the new exponent ('conformal spin') $\eta$  
parametrizes the algebra of operator product expansion. In the case of interest, 
the latter results from fusing two single fermions into a composite boson 
representing the Copper pairs, so that Eq.(63) then mimics the linearized gap equation 
(30) derived in the ME approximation.  

The difference with the extensively studied OTOC problem is that 
instead of the antisymmetric (odd under time reversal) eigenvectors
$\Psi^{-}_{\eta}$ of the kernal (64) the Cooper pairing problem focuses on the 
symmetric ($\Psi^{+}_{\eta}$ or even-frequency) ones.  
The latter can then be identified as a two-fermion wave function.

Per the above discussion, the structure of the gap equation crucially depends 
on whether the boson interaction function is evaluated in the weak or strong coupling regime. 
In the former, the functional form of the 
kernel in the gap equation readily conforms to Eq.(30) 
where the frequency integral 
becomes logarithmic for a constant $\Phi(\omega)$ and $\Delta_{\psi}=1/q$, 
thus reproducing the standard BCS scenario.
By contrast, in the latter regime it behaves  as a logarithm for $q=4$ only. 

Computing the integral in (63) with the ansatz (65) by using the formulas in the Appendix  
one obtains the eigenvalue 
\bea
\lambda(\Delta_{\psi},\eta)=-(\sin(2\pi\Delta_{\psi})+\sin(\pi\eta))\nonumber\\
{\Gamma(1-2\Delta_{\psi})\Gamma(2\Delta_{\psi}-\eta)\Gamma(2\Delta_{\psi}+\eta-1) 
\over {\pi\Gamma(2\Delta_{\psi}-1)}}
\label{eq:lambda-general}
\eea
an equivalent form of which 
was also obtained in the recent concurrent work \cite{schmalian_last}
after the release of the original version of the current manuscript.

Notably, the convergence of the integrals in Eq.(63) requires 
$1-2\Delta_{\psi}< Re\eta < 2\Delta_{\psi}$, thus narrowing the range of both parameters. 
For $q\to\infty$ (or $\Delta_{\psi}\to 0$ which is a hallmark of the identity operator) 
Eq.(66) yields $\lambda\approx 2\Delta_{\psi}/\eta(1-\eta)\to 0$,
while for $\Delta_{\psi}=1/2$ (free fermion limit) one 
finds $\lambda=1$ for all $\eta$. Also,   
$\lambda=-1$ for, both, $\eta=0$ and $\eta=1$ and all $\Delta_{\psi}$,
while for $\Delta_{\psi}=1/4$ (SYK fermion) 
the eigenvalue (66) reduces to 
$
\lambda(1/4,\eta)=
{1+\sin(\pi\eta)/{(2\eta-1)\cos(\pi\eta)}}
$.

Furthermore, $\lambda$ diverges at $\eta=2n+2\Delta_{\psi}$ 
and at $\eta=1-2n-2\Delta_{\psi}$ for all non-negative 
integer $n$, such poles corresponding to
the two-fermion composite operators (Cooper bound states).
In turn, the dimensions of the singular modes which signal the onset of the IR instabilities are
determined by the equation
\be 
\lambda(\Delta_{\psi},\eta)=1
\ee
In most of the previously studied OTOC problems, there exists 
the antisymmetric mode with the dimension $\eta=2$ 
for any $\Delta_{\psi}$. It is customarily associated with  
the emergent (pseudo-)Goldstone mode ('boundary graviton')
governed by the so-called Schwarzian action,  
all the other discrete roots being irrational, $\eta\approx 3.77, 5.68, 7.63,\dots$
\cite{Kitaev2015,kitaev_notes_2018,kitaev_soft_2018,kitaev_statistical_2019,sachdev_holographic_2010,sachdev_bekenstein-hawking_2015,sachdev_strange_2023,maldacena_bound_2016,maldacena_remarks_2016,maldacena_conformal_2016,stanford_fermionic_2017,polchinski_spectrum_2016,gross_generalization_2017,gross_all_2017,sarosi_ads_2_2018,bagrets_sachdev-ye-kitaev_2016,bagrets_power-law_2017,mertens_solving_2017,mertens_schwarzian_2018,yang_quantum_2019,jha_theory_2026,
khveshchenko_thickening_2018,khveshchenko_seeking_2018,khveshchenko_true_2019,khveshchenko_one_2020,
khveshchenko_gloria_2022}.

At first sight, the eigenvalue spectrum of the symmetric  
solutions $\Psi^{+}_{\eta}$ appears to be similar,
consisting of the discrete and continuum series. 
However, in contrast to the OTOC functions,  
the only viable solutions of Eq.(63) appear to be those of the continuum. 

Indeed,  the discrete series of eigenvalues $\eta=2n+2$ where $n\in N$  
is given by the expression 
$\lambda(\Delta_{\psi},2+2n)=-(2\Delta_{\psi})_{2n+1}/
(2\Delta_{\psi}-2n-2)_{2n+2}$ which is written in terms of 
the Pochhammer index $x_k=\Gamma(x+k)/\Gamma(x)$ and confined to the interval
$[0,1]$. In particular, for $n=0$ one obtains 
$\lambda(\Delta_{\psi},2)=\Delta_{\psi}/(1-\Delta_{\psi})$, which
falls below the minimal threshold (67) for all $\Delta_{\psi}<1/2$
(for instance, $\lambda(1/4,2)=1/3$). 

In contrast, evaluating (66) for the continuous series 
$\eta={1/2}+is$ where $s>0$ one obtains
\bea 
\lambda(\Delta_{\psi},1/2+is)={1-2\Delta_{\psi}\over \sin(2\pi\Delta_{\psi})}
|\Gamma(2\Delta_{\psi}-1/2+is)|^2\nonumber\\
(\cosh(\pi s)+\sin(2\pi\Delta_{\psi}))~~~~~~~~~~
\eea
In particular, $\lambda(1/4,1/2+is)={\coth(\pi s/2)/2s}$
which takes arbitrarily large values by diverging as $1/s^2$
for $s\to 0$. This behavior allows the eigenvalue $\lambda(\Delta_{\psi},1/2+is)$
 to readily attain 
the deep pairing regime ($\lambda>1$) at small $s$, while 
its $1/s$ decay (hence, $\lambda<1$) for $s\to\infty$ guarantees a necessary
crossing of the $\lambda=1$ threshold at some intermediate value of $s$
for all $\Delta_{\psi}<1/2$.  

Thus, putting $\eta=1/2$ in (68) and plugging it into Eq.(67) 
yields the dimension(s) $\Delta_{\psi}$ at  
which the corresponding eigenvector from the continuum goes critical.
Notably, the resulting equation has the reflection 
symmetry $\Delta_{\psi}\to 1-\Delta_{\psi}$ and, therefore, two solutions:
$\Delta_{\psi,1}\approx 0.07$ and $\Delta_{\psi,2}\approx 0.93$. 

Identifying $s$ with ${\sqrt {g-g_c}}$ one can recognize the above continuous values of the 
exponent $\eta$ as those given by Eq.(32). 
This analysis of the solutions of the gap equation (63) partially confirms
the previously predicted behavior of the pairing susceptibility that was argued to  
not change its frequency dependence (let alone, diverge) upon crossing the superconducting transition \cite{pandey_ward_2024,liu_towards_2021,palle_comment_2024,liu_reply_2024,abanov_non-bcs_2024,gindikin_collective_2025}. 

This would indeed be true for the contributions associated with non-critical modes 
whose (partial) pairing susceptibilities behave as $\chi_{\eta}\sim 1/(1-\lambda(\Delta_{\psi},\eta))$ and, therefore, remain finite 
for the generic $\Delta_{\psi}$ and $\eta$. However, once any of those modes becomes critical,  
the overall susceptibility $\sum_{\eta}\chi_{\eta}$ develops a divergence, 
provided that the term in question does not have a vanishing residue ('skipping pole').

In the presence of conserved supercharges, the eigenvalue $\lambda=1$ still corresponds to the bound two-fermion soft (quasi-zero) modes. However, similar BS equations can be formulated for the 
two-boson ($<\phi\phi>$) as well as the mixed ($<\phi\psi>$) two-point functions
describing bound states in the corresponding channels. 
Specifically, in the case of unbroken ${\cal N}=1$ or $2$ SUSY the bosonic field of dimension $\eta_b=2$ forms a supermultiplet with the fermion of dimension $\eta_f=3/2$ 
in, both, the even and odd channels. 
Moreover, for ${\cal N}=4$ this multiplet also includes another boson of dimension $\eta_b=1$  \cite{fu_supersymmetric_2017,peng_soft_2021,biggs_supersymmetric_2024,heydeman_phases_2023,gates_1d_2021,peng_correlators_2017,marcus_new_2019,heydeman_supersymmetry_2024,murugan_more_2017,benini_mathcaln2_2025,berkooz_double_2020}.

Furthermore, breaking $N=2$ SUSY down to ${\cal N}=1$ may result in the emergence of non-trivial 
correlators $G_{\psi\psi}$ and $G_{\phi\phi}$, in addition to
$G_{\psi{\bar \psi}}$ and $G_{\phi{\phi^{*}}}$. 
In particular, the amplitude $<\psi{\bar \psi}>$ stands for the particle-hole  
bound state which, upon condensing, becomes a CDW order parameter.
The problem of emergent CDW order 
can also be addressed by solving Eq.(63) with $\lambda(\Delta_{\psi},\eta)=-1$ where the sign difference originates from comparing the ladders in the two-fermion and fermion-hole channels, the two being related through the substitution $G(\tau)\to G(-\tau)=-G(\tau)$. 
\\

\noindent
{\it Holographic mirages}
\\

\noindent
As mentioned above, the zero-spin boson with $\eta_b=2$ corresponds to the (pseudo-)Goldstone soft mode which breaks reparametrization invariance, both, spontaneously and explicitly. 
In the IR limit, it is described by the 'gravitational' Schwarzian action which is  native to the SYK model, including its SUSY variants. 

Despite the popular claim, though, the underlying $AdS_2/CFT_1$ relationship
does not quite rise to the level of genuine holographic correspondence, as envisioned in Refs. \cite{hartnoll_lectures_2009,mcgreevy_holographic_2010,sachdev_what_2012,zaanen_holographic_2015,ammon_gaugegravity_2015}. Indeed, the bulk $1+1$-dimensional JT gravity is topological and, therefore, lacks any bulk dynamics of its own, being fully determined by the behavior on its $0+1$-dimensional boundary.
Moreover, this also holds true in the $2+1$ dimensions - which relationship serves as the basis for the $AdS_3/CFT_2$ duality \cite{jevicki_bi-local_2016,jevicki_bi-local_2016-1,das_three_2017,das_space-time_2018,mandal_coadjoint_2017,gaikwad_holographic_2020,mertens_defects_2019}. 

While the (yet to be fully ascertained) exact $AdS_2/CFT_1$ dual of the SYK model is expected to have an infinite tower of additional massive bulk 'matter' fields corresponding to the higher-energy states
(cf. the above discrete series of eigenstates in the pairing problem), thus far all the practical uses of this (pseudo-)duality have been almost entirely limited to the low-energy Schwarzian sector \cite{Kitaev2015,kitaev_notes_2018,kitaev_soft_2018,kitaev_statistical_2019,sachdev_holographic_2010,sachdev_bekenstein-hawking_2015,sachdev_strange_2023,maldacena_bound_2016,maldacena_remarks_2016,maldacena_conformal_2016,stanford_fermionic_2017,polchinski_spectrum_2016,gross_generalization_2017,gross_all_2017,sarosi_ads_2_2018,bagrets_sachdev-ye-kitaev_2016,bagrets_power-law_2017,mertens_solving_2017,mertens_schwarzian_2018,yang_quantum_2019,jha_theory_2026,khveshchenko_thickening_2018,khveshchenko_seeking_2018,khveshchenko_true_2019,khveshchenko_one_2020,khveshchenko_gloria_2022}.

In contrast, 
the original applied holographic conjecture posits that such a correspondence should involve pairs of systems operating in genuinely different dimensions \cite{hartnoll_lectures_2009,mcgreevy_holographic_2010,sachdev_what_2012,zaanen_holographic_2015,ammon_gaugegravity_2015}. 
In that regard, in order to make a strong case for holography one would better strive to identify and 
juxtapose those pairs of systems that demonstrate identical physical properties while showing no obvious sign of a possible reduction  to the same mathematical description in the lesser of the two dimensions. 

Conversely, the very possibility of such a reduction  would indicate that, despite their different formal appearance, both systems belong in the 
same space-time dimension, thus revealing that the higher-dimensional (bulk) description is, in fact, redundant and fully reducible to the lower-dimensional (boundary) one. 

Also, in contrast with the general philosophy of the hypothetical generalized ('$non-AdS/non-CFT$') holographical conjecture \cite{hartnoll_lectures_2009,mcgreevy_holographic_2010,sachdev_what_2012,zaanen_holographic_2015,ammon_gaugegravity_2015}, the $AdS_2/CFT_1$ duality does not offer any immediate way of relating a weak-coupling regime 
on one side of the purported correspondence to a strong one on the 
other (also, for the sheer lack of either), thereby largely negating any practical purpose of such would-be holographic construction.

Thus, when taken at its face value, the purported holographic duality amounts to little more than the Stokes' formulas for the boundary value problem in classical electromagnetism where
in the absence of bulk charges the fields lack any independent dynamics of their own. 
Instead of providing an iron-clad showcase of genuine inter-dimensional correspondence, it then constitutes 
a mere example of 'Hall-o-graphy', as per its definition in Refs.\cite{khveshchenko_diagnostic_2025,khveshchenko_it_2023,khveshchenko_phase_2022}.

The connotation for the Hall effect then follows from the effective dimensional reduction, akin to that (from two down to one) of the phase space dimension in the strong-field Hall effect when all the electrons are confined to the lowest Landau level. 
The possibility of this natural lowering of the dimension hinges on the topological nature of the setup in question - which mechanism would indeed be expected to work 
in $AdS_2/CFT_1$ (as well as $AdS_3/CFT_2$), but not in a generic spatial dimension $d>1$.
 
Yet another limitation of the (pseudo-)holographic $AdS_2/CFT_1$ relationship is that the $AdS_2$ geometry appears to describe the asymptotic behavior of the entire host of 
higher-dimensional 'near-critical' geometries of the direct product type
$AdS_2\otimes R^{d}$,
including the physically relevant $d=2$- and $3$-dimensional ones. Thus, it reduces the multitude of the bulk metrics (Reissner-Nordstrom, Gubser-Rocha, Lifshitz, hyperscaling-violating, Bianchi, Q-lattices, etc.) 
that have been opportunistically introduced 'ad hoc'  and 
claimed to provide viable descriptions of the various concrete physical systems \cite{hartnoll_lectures_2009,mcgreevy_holographic_2010,sachdev_what_2012,zaanen_holographic_2015,ammon_gaugegravity_2015}, to just a single one.

On the positive side, though, the IR Schwarzian description provides a unifying framework for  the universality class characterized by 'ultra-locality' in the coordinate space, yet extended temporal correlations. In the Refs.\cite{hartnoll_lectures_2009,mcgreevy_holographic_2010,sachdev_what_2012,zaanen_holographic_2015,ammon_gaugegravity_2015}, 
this regime would often be referred to as that of the dynamical critical index $z=\infty$, 
which behavior bears a strong resemblance to the popular conventional approach dubbed dynamical mean-field theory. 

Such affinity between the two approaches has already helped one to bring the generalized holographic conjecture \cite{hartnoll_lectures_2009,mcgreevy_holographic_2010,sachdev_what_2012,zaanen_holographic_2015,ammon_gaugegravity_2015} closer to the mainstream condensed matter theory, thus potentially offering some (perhaps, anecdotal) support for 
the former's being 'possibly right, albeit for the wrong reason' \cite{khveshchenko_diagnostic_2025,khveshchenko_it_2023,khveshchenko_phase_2022}. 

In the specific context of the YSYK pairing problem, some additional arguments in favor the holographic connection were put forward as well. Specifically, it was pointed out that introducing 
the so-called Radon transform of the pairing field 
\be
\Psi(T,\epsilon)=\epsilon^{(\gamma-1)/2}\oint_{\Gamma}\Phi(\tau,\zeta)dl
\label{eq:radon}
\ee
where the contour $\Gamma$ in the $(\tau,\zeta)$-plane is defined by the equation $(\tau-T)^2+\zeta^2=1/\epsilon^2$,
gives rise to the emergent $AdS_2$ metric (and its thermal counterpart for $T>0$) \cite{davison_thermoelectric_2017,cheipesh_planckian_2019,inkof_quantum_2022,esterlis_quantum_2025,shankar_josephson_2025,stone_radon_2025}.

It appears, however, that this observation would be common to the generic bi-local observables 
whose $AdS_2$ kinematic space is constructed with the use of 
the light-cone coordinates $(\tau,\zeta)$ \cite{jevicki_bi-local_2016,jevicki_bi-local_2016-1,das_three_2017,das_space-time_2018,mandal_coadjoint_2017,gaikwad_holographic_2020,mertens_defects_2019}. 
Moreover, unlike the aforementioned $AdS_2\otimes R^d$-based holographic scenarios, such a construction is specifically $1+1$-dimensional 
and does not allow for an immediate generalization to any higher spatial dimension $d>1$.

Lastly, as mentioned above, a practical holographic scheme would normally be supposed to relate   
a strongly/weakly coupled boundary description to a weakly/strongly coupled bulk one,
a formal implementation of which relationship would necessarily involve a $non-linear$ field transformation.
Instead, using the $linear$ (albeit non-local) transform (68) could hardly succeed in linearizing any    
truly non-linear problem, thus making questionable the usefulness of such transformation for any practical calculation.
Therefore, it should come as no surprise that the bulk dual theory appears to 
be that of a weakly-interacting (or even free) massive boson in the $AdS_2$ space-time \cite{davison_thermoelectric_2017,cheipesh_planckian_2019,inkof_quantum_2022,esterlis_quantum_2025,shankar_josephson_2025,stone_radon_2025}. 
\\  

\noindent
{\it Summary}
\\

\noindent
The various SUSY variants of the popular (Y)SYK model provide another step towards a better 
understanding of the narrow-band strongly correlated systems. 
They feature exact power-law solutions in the regime where superconductivity coexists with 
(and emerges from) the NFL normal states, thus offering additional means of checking on the applicability of such customary approximations as the ME one.  

In this work, we exposed a number of subtleties associated with the customary use of 
the ME approximation. Specifically, this analysis  
emphasizes the need to further explore the limits of applicability of 
the ME theory in the variants of the (Y)SYK model, 
with the focus on the properties of the actual materials which those models seek to represent. 

To that end, the SUSY YSYK models yields exact anomalous dimensions, including $rational$ ones (see, e.g., (49) and (62)). In contrast, the previous studies focused exclusively on the $irrational$ (non-SUSY) solutions. 
It is worth noting that the self-consistent equation (48) determining such solutions seems to have been 
notoriously difficult to reproduce in the error-free form. In particular, it was presented incorrectly in the recent review article \cite{abchub} that appeared after the release of the original version of the current manuscript.  
 
Ascertaining the status of the ME approximation and the results obtained within its framework  
can help one to evaluate the previous reports
of the unusual behavior of the dynamical pairing susceptibility which, contrary to what happens in the conventional BCS scenario, was claimed to neither diverge at the onset of pairing, nor change 
its functional form \cite{abanov_interplay_2020,wu_interplay_2020,wu_interplay_2020-1,wu_interplay_2021,wu_interplay_2021-1,
zhang_interplay_2021,yu_topologically_2026,pandey_ward_2024,liu_towards_2021,palle_comment_2024,liu_reply_2024,abanov_non-bcs_2024,gindikin_collective_2025}.
In Refs.\cite{pandey_ward_2024,liu_towards_2021,palle_comment_2024,liu_reply_2024,abanov_non-bcs_2024,gindikin_collective_2025} it was conjectured that such behavior of the
susceptibility may reflect a multi-critical nature of the superconducting transition, as  
an infinite tower of superconducting states described by the oscillatory (non-monotonic) 
solutions with different numbers of nodes on the frequency axis emerges simultaneously. 

To that end, we performed a systematic study of the even-frequency pairing problem 
by finding the proper (symmetric, as opposed to the previously studied 
antisymmetric, see the literature on the OTOC problem [11-18])
asymptotically exact large-$N$ solutions of the 
underlying integral gap equation (30), an equivalent form of which was later   
presented in the concurrent Ref.\cite{schmalian_last}).

Such solutions appear to be parametrized by a continuously-varying exponent $\eta=1/2+is$ where $s={\sqrt {g-g_0}}>0$ measures the excess of the electron-phonon coupling over its critical value.
Applying the boundary conditions on the gap function then  
yields a sequence of the progressively decreasing critical 
temperatures $T_n\sim\Lambda e^{-\pi(n+1/2)/s}$
where $\Lambda$ stands for the UV cutoff 
and $n$ is a non-negative integer (cf. Eq.(34)). 

Upon cranking up the coupling strength $g$ the soft modes described by such solutions pass one-by-one 
through the singular point one where their pairing susceptibilities diverge, while the rest of the spectrum remains non-critical. 
Nevertheless, akin to the previously studied examples of such behavior [103-108], 
the $n=0$ solution corresponding to the monotonically decreasing gap function (33) 
sets in first, thereby preempting any further instabilities associated with the non-monotonic 
(for $n>0$) gap functions
$\Delta_n(\omega)\sim\sin(\pi (n+1/2){\ln(\Lambda/\omega)/\ln(\Lambda/T_n)})$.   

The recent developments (which remain partly in disagreement)  \cite{tsoncheva_condensation_2005,marsiglio_electron_2001,mirabi_thermodynamics_2020,zhang_superconductivity_2022,zhang_free_2023-1,paci_polaronic_2005,fratini_optical_2006,heidarian_inhomogeneous_2003,wolfle_self-consistent_2010,miskic_polaron_2025,ragni_polarons_2025,chubukov_breakdown_2026,prokofev_limits_2026,chubukov_eliashberg_2020,cai_superconductivity_2025,kabanov_adiabatic_2020,yuzbashyan_superconductivity_2022,yuzbashyan_breakdown_2022,yuzbashyan_migdal-eliashberg_2022,yuzbashyan_instability_2025,kiessling_bounds_2025,kiessling_bounds_2025-1,kiessling_bounds_2025-2,semenok_fundamental_2025,abanov_interplay_2020,wu_interplay_2020,wu_interplay_2020-1,wu_interplay_2021,wu_interplay_2021-1,zhang_interplay_2021,yu_topologically_2026} show that the continuing saga of the classical electron-phonon problem remains far from over.
In that regard, the (Y)SYK and related (asymptotically) solvable models provide an opportunity to study this and related problems in a controlled setting and without relying on some popular, yet potentially limiting approximations.
\\

\noindent
{\it Acknowledgments}
\\

The author acknowledges hospitality at the Aspen Center for Physics, 
International Institute of Physics in Natal, ETH Zurich, and  
Institute of Laue-Langevin in Grenoble which were visited during this work. 
\\

\noindent
{\it Appendix}
\\

\noindent
\be
\int dx {e^{ixy}\over |x|^{\alpha}}=2{\Gamma(1-\alpha)\cos(\pi \alpha/2)|y|^{\alpha-1}}
\label{eq:app-fourier-even}
\ee

\be
\int sgn x dx {e^{ixy}\over |x|^{\alpha}}=2i sgn y{\Gamma(1-\alpha)\sin(\pi \alpha/2)|y|^{\alpha-1}}
\label{eq:app-fourier-odd}
\ee

\be
\int {dx\over |x|^{\alpha}|x+y|^{\beta}}=
{{\sqrt \pi}\Gamma({1-\alpha\over 2})\Gamma({1-\beta\over 2})\Gamma({\alpha+\beta-1\over 2})
\over
\Gamma({\alpha\over 2})\Gamma({\beta\over 2})\Gamma({1-{\alpha+\beta\over 2})}
|y|^{\alpha+\beta-1}}
\label{eq:app-conv-even-even}
\ee

\be
\int {sgn x sgn(x+y) dx\over |x|^{\alpha}|x+y|^{\beta}}={I^{+}_{\alpha\beta}\over |y|^{\alpha+\beta-1}}
\label{eq:app-conv-even}
\ee

\be
\int {sgn (x+y) dx\over {|x|^{\alpha}|x+y|^{\beta}}}=
sgn y{I^{-}_{\alpha\beta}\over |y|^{\alpha+\beta-1}}
\label{eq:app-conv-odd}
\ee
where
\be
I^{\pm}_{\alpha\beta}=\beta(1-\alpha,\alpha+\beta-1)\pm \beta(1-\beta,\alpha+\beta-1)-\beta(1-\alpha,1-\beta)
\label{eq:app-I-coeff}
\ee 
and $\beta(\alpha,\beta)=\Gamma(\alpha)\Gamma(\beta)/\Gamma(\alpha+\beta)$.

\bea
\int {sgn x sgn y sgn (y-x+z) dxdy\over {|x|^{2\alpha}|y|^{2\beta}|y-x+z|^{2\gamma}}}=~~~~~~~~~~~\\
{\pi^2sgn z\over |z|^{2\alpha+2\beta+2\gamma-2}} 
{\Gamma(2\alpha+2\beta+2\gamma-2)\sin\pi(\alpha+\beta+\gamma-1) \over
{\Gamma(2\alpha)\Gamma(2\beta)\Gamma(2\gamma)\sin\pi\alpha\sin\pi\beta\sin\pi\gamma}}\nonumber
\label{eq:app-triple-sss}
\eea

\bea
\int {sgn x sgn y dxdy\over {|x|^{2\alpha}|y|^{2\beta}|y-x+z|^{2\gamma}}}=~~~~~~~~~~\\
{\pi^2\over |z|^{2\alpha+2\beta+2\gamma-2}} 
{\Gamma(2\alpha+2\beta+2\gamma-2)\cos\pi(\alpha+\beta+\gamma-1) \over
{\Gamma(2\alpha)\Gamma(2\beta)\Gamma(2\gamma)\sin\pi\alpha\sin\pi\beta\cos\pi\gamma}}\nonumber
\label{eq:app-triple-ssn}
\eea

\bea
\int {sgn x sgn (y-x+z) dxdy\over {|x|^{2\alpha}|y|^{2\beta}|y-x+z|^{2\gamma}}}=~~~~~~~~~~\\
-{\pi^2\over |z|^{2\alpha+2\beta+2\gamma-2}} 
{\Gamma(2\alpha+2\beta+2\gamma-2)\cos\pi(\alpha+\beta+\gamma-1) \over
{\Gamma(2\alpha)\Gamma(2\beta)\Gamma(2\gamma)\sin\pi\alpha\cos\pi\beta\sin\pi\gamma}}\nonumber
\label{eq:app-triple-sns}
\eea

\bea
\int {sgn x dxdy\over {|x|^{2\alpha}|y|^{2\beta}|y-x+z|^{2\gamma}}}=~~~~~~~~~~\\
{\pi^2sgn z\over |z|^{2\alpha+2\beta+2\gamma-2}} 
{\Gamma(2\alpha+2\beta+2\gamma-2)\sin\pi(\alpha+\beta+\gamma-1) \over
{\Gamma(2\alpha)\Gamma(2\beta)\Gamma(2\gamma)\sin\pi\alpha\cos\pi\beta\cos\pi\gamma}}\nonumber
\label{eq:app-triple-snn}
\eea

\bea
\int {sgn (y-x+z) dxdy\over {|x|^{2\alpha}|y|^{2\beta}|y-x+z|^{2\gamma}}}=~~~~~~~~~~\\
{\pi^2sgn z\over |z|^{2\alpha+2\beta+2\gamma-2}} 
{\Gamma(2\alpha+2\beta+2\gamma-2)\sin\pi(\alpha+\beta+\gamma-1) \over
{\Gamma(2\alpha)\Gamma(2\beta)\Gamma(2\gamma)\cos\pi\alpha\cos\pi\beta\sin\pi\gamma}}\nonumber
\label{eq:app-triple-nns}
\eea

\bea
\int {dxdy\over {|x|^{2\alpha}|y|^{2\beta}|y-x+z|^{2\gamma}}}=~~~~~~~~~~\\
{\pi^2\over |z|^{2\alpha+2\beta+2\gamma-2}} 
{\Gamma(2\alpha+2\beta+2\gamma-2)\cos\pi(\alpha+\beta+\gamma-1) \over
{\Gamma(2\alpha)\Gamma(2\beta)\Gamma(2\gamma)\cos\pi\alpha\cos\pi\beta\cos\pi\gamma}}\nonumber
\label{eq:app-triple-nnn}
\eea

\newpage

\bibliographystyle{apsrev4-2}
\bibliography{references}

@article{french_random-matrix_1971,
	title = {Some random-matrix level and spacing distributions for fixed-particle-rank interactions},
	volume = {35},
	copyright = {https://www.elsevier.com/tdm/userlicense/1.0/},
	issn = {03702693},
	url = {https://linkinghub.elsevier.com/retrieve/pii/0370269371904242},
	doi = {10.1016/0370-2693(71)90424-2},
	number = {1},
	urldate = {2026-07-17},
	journal = {Physics Letters B},
	author = {French, J.B. and Wong, S.S.M.},
	month = apr,
	year = {1971},
	pages = {5--7},
}

@article{bohigas_spacing_1971,
	title = {Spacing and individual eigenvalue distributions of two-body random hamiltonians},
	volume = {35},
	copyright = {https://www.elsevier.com/tdm/userlicense/1.0/},
	issn = {03702693},
	url = {https://linkinghub.elsevier.com/retrieve/pii/0370269371903996},
	doi = {10.1016/0370-2693(71)90399-6},
	number = {5},
	urldate = {2026-07-17},
	journal = {Physics Letters B},
	author = {Bohigas, O. and Flores, J.},
	month = jun,
	year = {1971},
	pages = {383--386},
}

@article{sachdev_gapless_1993,
	title = {Gapless {Spin}-{Fluid} {Ground} {State} in a {Random} {Quantum} {Heisenberg} {Magnet}},
	volume = {70},
	issn = {0031-9007},
	url = {http://arxiv.org/abs/cond-mat/9212030},
	doi = {10.1103/PhysRevLett.70.3339},
	number = {21},
	urldate = {2026-07-17},
	journal = {Physical Review Letters},
	author = {Sachdev, Subir and Ye, Jinwu},
	month = may,
	year = {1993},
	note = {arXiv:cond-mat/9212030},
	pages = {3339--3342},
}

@misc{Kitaev2015,
  author       = {Kitaev, Alexei},
  title        = {A simple model of quantum holography},
  howpublished = {Talks at KITP, University of California, Santa Barbara},
  note         = {Part 1: 7 April 2015; Part 2: 27 May 2015.
                  http://online.kitp.ucsb.edu/online/entangled15/kitaev/},
  year         = {2015}
}

@misc{kitaev_notes_2018,
	title = {Notes on {$\widetilde{SL}(2,\mathbb{R})$} representations},
	url = {http://arxiv.org/abs/1711.08169},
	doi = {10.48550/arXiv.1711.08169},
	urldate = {2026-07-17},
	publisher = {arXiv},
	author = {Kitaev, Alexei},
	month = aug,
	year = {2018},
	note = {arXiv:1711.08169 [hep-th]},
}

@article{kitaev_soft_2018,
	title = {The soft mode in the {Sachdev}-{Ye}-{Kitaev} model and its gravity dual},
	volume = {2018},
	issn = {1029-8479},
	url = {http://arxiv.org/abs/1711.08467},
	doi = {10.1007/JHEP05(2018)183},
	number = {5},
	urldate = {2026-07-17},
	journal = {Journal of High Energy Physics},
	author = {Kitaev, Alexei and Suh, S. Josephine},
	month = may,
	year = {2018},
	note = {arXiv:1711.08467 [hep-th]},
	pages = {183},
}

@article{kitaev_statistical_2019,
	title = {Statistical mechanics of a two-dimensional black hole},
	volume = {2019},
	issn = {1029-8479},
	url = {http://arxiv.org/abs/1808.07032},
	doi = {10.1007/JHEP05(2019)198},
	number = {5},
	urldate = {2026-07-17},
	journal = {Journal of High Energy Physics},
	author = {Kitaev, Alexei and Suh, S. Josephine},
	month = may,
	year = {2019},
	note = {arXiv:1808.07032 [hep-th]},
	pages = {198},
}

@article{sachdev_holographic_2010,
	title = {Holographic metals and the fractionalized {Fermi} liquid},
	volume = {105},
	issn = {0031-9007, 1079-7114},
	url = {http://arxiv.org/abs/1006.3794},
	doi = {10.1103/PhysRevLett.105.151602},
	number = {15},
	urldate = {2026-07-17},
	journal = {Physical Review Letters},
	author = {Sachdev, Subir},
	month = oct,
	year = {2010},
	note = {arXiv:1006.3794 [hep-th]},
	pages = {151602},
}

@article{sachdev_bekenstein-hawking_2015,
	title = {Bekenstein-{Hawking} {Entropy} and {Strange} {Metals}},
	volume = {5},
	issn = {2160-3308},
	url = {http://arxiv.org/abs/1506.05111},
	doi = {10.1103/PhysRevX.5.041025},
	number = {4},
	urldate = {2026-07-17},
	journal = {Physical Review X},
	author = {Sachdev, Subir},
	month = nov,
	year = {2015},
	note = {arXiv:1506.05111 [hep-th]},
	pages = {041025},
}

@incollection{sachdev_strange_2023,
	title = {Strange metals and black holes: insights from the {Sachdev}-{Ye}-{Kitaev} model},
    booktitle = {Oxford Research Encyclopedia of Physics},
	publisher = {Oxford University Press},
	shorttitle = {Strange metals and black holes},
	url = {http://arxiv.org/abs/2305.01001},
	doi = {10.1093/acrefore/9780190871994.013.48},
	urldate = {2026-07-17},
	author = {Sachdev, Subir},
	month = dec,
	year = {2023},
	note = {arXiv:2305.01001 [cond-mat.str-el]},
}

@article{maldacena_bound_2016,
	title = {A bound on chaos},
	volume = {2016},
	issn = {1029-8479},
	url = {http://arxiv.org/abs/1503.01409},
	doi = {10.1007/JHEP08(2016)106},
	number = {8},
	urldate = {2026-07-17},
	journal = {Journal of High Energy Physics},
	author = {Maldacena, Juan and Shenker, Stephen H. and Stanford, Douglas},
	month = aug,
	year = {2016},
	note = {arXiv:1503.01409 [hep-th]},
	pages = {106},
}

@article{maldacena_remarks_2016,
	title = {Remarks on the {Sachdev}-{Ye}-{Kitaev} model},
	volume = {94},
	doi = {10.1103/PhysRevD.94.106002},
	number = {10},
	journal = {Physical Review D},
	author = {Maldacena, Juan and Stanford, Douglas},
	year = {2016},
	pages = {106002},
}

@misc{maldacena_conformal_2016,
	title = {Conformal symmetry and its breaking in two dimensional {Nearly} {Anti}-de-{Sitter} space},
	url = {http://arxiv.org/abs/1606.01857},
	doi = {10.48550/arXiv.1606.01857},
	urldate = {2026-07-17},
	publisher = {arXiv},
	author = {Maldacena, Juan and Stanford, Douglas and Yang, Zhenbin},
	month = sep,
	year = {2016},
	note = {arXiv:1606.01857 [hep-th]},
}

@article{stanford_fermionic_2017,
	title = {Fermionic {Localization} of the {Schwarzian} {Theory}},
	volume = {2017},
	issn = {1029-8479},
	url = {http://arxiv.org/abs/1703.04612},
	doi = {10.1007/JHEP10(2017)008},
	number = {10},
	urldate = {2026-07-17},
	journal = {Journal of High Energy Physics},
	author = {Stanford, Douglas and Witten, Edward},
	month = oct,
	year = {2017},
	note = {arXiv:1703.04612 [hep-th]},
	pages = {8},
}

@article{polchinski_spectrum_2016,
	title = {The {Spectrum} in the {Sachdev}-{Ye}-{Kitaev} {Model}},
	volume = {2016},
	issn = {1029-8479},
	url = {http://arxiv.org/abs/1601.06768},
	doi = {10.1007/JHEP04(2016)001},
	number = {4},
	urldate = {2026-07-17},
	journal = {Journal of High Energy Physics},
	author = {Polchinski, Joseph and Rosenhaus, Vladimir},
	month = apr,
	year = {2016},
	note = {arXiv:1601.06768 [hep-th]},
	pages = {1--25},
}

@article{gross_generalization_2017,
	title = {A {Generalization} of {Sachdev}-{Ye}-{Kitaev}},
	volume = {2017},
	issn = {1029-8479},
	url = {http://arxiv.org/abs/1610.01569},
	doi = {10.1007/JHEP02(2017)093},
	number = {2},
	urldate = {2026-07-17},
	journal = {Journal of High Energy Physics},
	author = {Gross, David J. and Rosenhaus, Vladimir},
	month = feb,
	year = {2017},
	note = {arXiv:1610.01569 [hep-th]},
	pages = {93},
}

@article{gross_all_2017,
	title = {All point correlation functions in {SYK}},
	volume = {2017},
	issn = {1029-8479},
	url = {http://arxiv.org/abs/1710.08113},
	doi = {10.1007/JHEP12(2017)148},
	number = {12},
	urldate = {2026-07-17},
	journal = {Journal of High Energy Physics},
	author = {Gross, David J. and Rosenhaus, Vladimir},
	month = dec,
	year = {2017},
	note = {arXiv:1710.08113 [hep-th]},
	pages = {148},
}

@inproceedings{sarosi_ads_2_2018,
	title = {{AdS}{$_2$} holography and the {SYK} model},
	url = {http://arxiv.org/abs/1711.08482},
	doi = {10.22323/1.323.0001},
	urldate = {2026-07-17},
	booktitle = {Proceedings of {XIII} {Modave} {Summer} {School} in {Mathematical} {Physics} — {PoS}({Modave2017})},
	author = {Sárosi, Gábor},
	month = mar,
	year = {2018},
	note = {arXiv:1711.08482 [hep-th]},
	pages = {001},
}

@article{bagrets_sachdev-ye-kitaev_2016,
	title = {Sachdev-{Ye}-{Kitaev} {Model} as {Liouville} {Quantum} {Mechanics}},
	volume = {911},
	issn = {05503213},
	url = {http://arxiv.org/abs/1607.00694},
	doi = {10.1016/j.nuclphysb.2016.08.002},
	urldate = {2026-07-17},
	journal = {Nuclear Physics B},
	author = {Bagrets, Dmitry and Altland, Alexander and Kamenev, Alex},
	month = oct,
	year = {2016},
	note = {arXiv:1607.00694 [cond-mat.str-el]},
	pages = {191--205},
}

@article{bagrets_power-law_2017,
	title = {Power-law out of time order correlation functions in the {SYK} model},
	volume = {921},
	doi = {10.1016/j.nuclphysb.2017.06.012},
	journal = {Nucl. Phys. B},
	author = {Bagrets, Dmitry and Altland, Alexander and Kamenev, Alex},
	year = {2017},
	note = {arXiv:1702.08902},
	pages = {727--752},
}

@article{mertens_solving_2017,
	title = {Solving the {Schwarzian} via the conformal bootstrap},
	volume = {2017},
	issn = {1029-8479},
	url = {https://doi.org/10.1007/JHEP08(2017)136},
	doi = {10.1007/JHEP08(2017)136},
	number = {8},
	urldate = {2026-07-17},
	journal = {Journal of High Energy Physics},
	author = {Mertens, Thomas G. and Turiaci, Gustavo J. and Verlinde, Herman L.},
	month = aug,
	year = {2017},
	pages = {136},
}

@article{mertens_schwarzian_2018,
	title = {The {Schwarzian} theory — origins},
	volume = {2018},
	issn = {1029-8479},
	url = {https://doi.org/10.1007/JHEP05(2018)036},
	doi = {10.1007/JHEP05(2018)036},
	number = {5},
	urldate = {2026-07-17},
	journal = {Journal of High Energy Physics},
	author = {Mertens, Thomas G.},
	month = may,
	year = {2018},
	pages = {36},
}

@article{yang_quantum_2019,
	title = {The {Quantum} {Gravity} {Dynamics} of {Near} {Extremal} {Black} {Holes}},
	volume = {2019},
	issn = {1029-8479},
	url = {http://arxiv.org/abs/1809.08647},
	doi = {10.1007/JHEP05(2019)205},
	number = {5},
	urldate = {2026-07-17},
	journal = {Journal of High Energy Physics},
	author = {Yang, Zhenbin},
	month = may,
	year = {2019},
	note = {arXiv:1809.08647 [hep-th]},
	pages = {205},
}

@book{jha_theory_2026,
	title = {Theory of {Strongly} {Correlated} {Systems}: {An} {Introduction} to {Sachdev}-{Ye}-{Kitaev} {Model}},
	volume = {1051},
	shorttitle = {Theory of {Strongly} {Correlated} {Systems}},
	url = {http://arxiv.org/abs/2507.07195},
	doi = {10.1007/978-3-032-20910-8},
	urldate = {2026-07-17},
	author = {Jha, Rishabh},
    publisher = {Springer},
	year = {2026},
	note = {arXiv:2507.07195 [hep-th]},
}

@article{khveshchenko_thickening_2018,
	title = {Thickening and sickening the {SYK} model},
	volume = {5},
	issn = {2542-4653},
	url = {http://arxiv.org/abs/1705.03956},
	doi = {10.21468/SciPostPhys.5.1.012},
	number = {1},
	urldate = {2026-07-17},
	journal = {SciPost Physics},
	author = {Khveshchenko, D. V.},
	month = jul,
	year = {2018},
	note = {arXiv:1705.03956 [cond-mat.str-el]},
	pages = {012},
}

@misc{khveshchenko_seeking_2018,
	title = {Seeking to develop global {SYK}-ness},
	url = {http://arxiv.org/abs/1805.00870},
	doi = {10.48550/arXiv.1805.00870},
	urldate = {2026-07-17},
	publisher = {arXiv},
	author = {Khveshchenko, D. V.},
	month = nov,
	year = {2018},
	note = {arXiv:1805.00870 [cond-mat.str-el]},
}

@article{khveshchenko_true_2019,
	title = {True {SYK} or (con)sequences},
	volume = {59},
	issn = {2424-3647, 1648-8504},
	url = {http://arxiv.org/abs/1905.04381},
	doi = {10.3952/physics.v59i2.4013},
	number = {2},
	urldate = {2026-07-17},
	journal = {Lithuanian Journal of Physics},
	author = {Khveshchenko, D. V.},
	month = aug,
	year = {2019},
	note = {arXiv:1905.04381 [hep-th]},
    pages = {104},
}

@article{khveshchenko_one_2020,
	title = {One {SYK} single electron transistor},
	volume = {60},
	issn = {2424-3647, 1648-8504},
	url = {http://arxiv.org/abs/1912.05691},
	doi = {10.3952/physics.v60i3.4305},
	number = {3},
	urldate = {2026-07-17},
	journal = {Lithuanian Journal of Physics},
	author = {Khveshchenko, D. V.},
	month = aug,
	year = {2020},
	note = {arXiv:1912.05691 [cond-mat.str-el]},
    pages = {185},
}

@article{khveshchenko_gloria_2022,
	title = {The gloria mundi of {SYK} does not transit yet},
	doi = {10.3952/physics.v62i2.4741},
	author = {Khveshchenko, D. V.},
    journal = {Lithuanian Journal of Physics},
	month = may,
	year = {2022},
	note = {arXiv:2205.11478},
    volume = {62},
    number = {2},
    pages = {81}
}

@misc{kleger_universal_2026,
	title = {Universal {Theory} of {Incoherent} {Metals}},
	url = {http://arxiv.org/abs/2605.03013},
	doi = {10.48550/arXiv.2605.03013},
	urldate = {2026-07-17},
	publisher = {arXiv},
	author = {Kleger, Aaron and Gnezdilov, Nikolay and Boyack, Rufus},
	month = may,
	year = {2026},
	note = {arXiv:2605.03013 [cond-mat.str-el]},
}

@article{bonetti_fractionalized_2026,
	title = {Fractionalized {Fermi} liquids and the cuprate phase diagram},
	volume = {89},
	issn = {0034-4885, 1361-6633},
	url = {http://arxiv.org/abs/2508.20164},
	doi = {10.1088/1361-6633/ae530d},
	number = {4},
	urldate = {2026-07-17},
	journal = {Reports on Progress in Physics},
	author = {Bonetti, Pietro M. and Christos, Maine and Nikolaenko, Alexander and Patel, Aavishkar A. and Sachdev, Subir},
	month = apr,
	year = {2026},
	note = {arXiv:2508.20164 [cond-mat.str-el]},
	pages = {044501},
}

@misc{sutradhar_singlet_2024,
	title = {Singlet, triplet, and mixed all-to-all pairing states emerging from incoherent fermions},
	url = {http://arxiv.org/abs/2404.03731},
	doi = {10.48550/arXiv.2404.03731},
	urldate = {2026-07-17},
	publisher = {arXiv},
	author = {Sutradhar, Jagannath and Ruhman, Jonathan and Klein, Avraham},
	month = apr,
	year = {2024},
	note = {arXiv:2404.03731 [cond-mat.str-el]},
}

@article{gnezdilov_upper_2025,
	title = {Upper bound on {$T_c$} in a strongly coupled electron-boson superconductor},
	volume = {112},
	issn = {2469-9950, 2469-9969},
	url = {http://arxiv.org/abs/2505.02894},
	doi = {10.1103/7zmy-xtcw},
	number = {18},
	urldate = {2026-07-17},
	journal = {Physical Review B},
	author = {Gnezdilov, Nikolay V. and Boyack, Rufus},
	month = nov,
	year = {2025},
	note = {arXiv:2505.02894 [cond-mat.str-el]},
	pages = {L180504},
}

@misc{elezaby_superconductivity_2026,
	title = {Superconductivity {Near} a {Quantum} {Critical} {Point}: {Bounds} on the {Transition} {Temperature} in the {$\gamma$}-{Model}},
	shorttitle = {Superconductivity {Near} a {Quantum} {Critical} {Point}},
	url = {http://arxiv.org/abs/2512.20009},
	doi = {10.48550/arXiv.2512.20009},
	urldate = {2026-07-17},
	publisher = {arXiv},
	author = {Elezaby, Ahmed and Abanov, Artem},
	month = jun,
	year = {2026},
	note = {arXiv:2512.20009 [cond-mat.supr-con]},
}

@article{choi_pairing_2022,
	title = {Pairing {Instabilities} of the {Yukawa}-{SYK} {Models} with {Controlled} {Fermion} {Incoherence}},
	volume = {12},
	issn = {2542-4653},
	url = {http://arxiv.org/abs/2110.02968},
	doi = {10.21468/SCIPOSTPHYS.12.5.151},
	number = {5},
	urldate = {2026-07-17},
	journal = {SciPost Physics},
	author = {Choi, Wonjune and Tavakol, Omid and Kim, Yong Baek},
	month = may,
	year = {2022},
	note = {arXiv:2110.02968 [cond-mat.str-el]},
	pages = {151},
}

@misc{zhang_density_2023,
	title = {Density of states and spectral function of a superconductor out of a quantum-critical metal},
	url = {http://arxiv.org/abs/2301.13679},
	doi = {10.48550/arXiv.2301.13679},
	urldate = {2026-07-17},
	publisher = {arXiv},
	author = {Zhang, Shang-Shun and Chubukov, Andrey V.},
	month = jan,
	year = {2023},
	note = {arXiv:2301.13679 [cond-mat.supr-con]},
}

@article{lantagne-hurtubise_superconducting_2021,
	title = {Superconducting instabilities in a spinful {Sachdev}-{Ye}-{Kitaev} model},
	volume = {104},
	url = {https://link.aps.org/doi/10.1103/PhysRevB.104.L020509},
	doi = {10.1103/PhysRevB.104.L020509},
	number = {2},
	urldate = {2026-07-17},
	journal = {Physical Review B},
	publisher = {American Physical Society},
	author = {Lantagne-Hurtubise, {\'E}tienne and Pathak, Vedangi and Sahoo, Sharmistha and Franz, Marcel},
	month = jul,
	year = {2021},
	pages = {L020509},
}

@article{valentinis_bcs_2023,
	title = {{BCS} to incoherent superconductivity crossovers in the {Yukawa}-{SYK} model on a lattice},
	volume = {108},
	doi = {10.1103/PhysRevB.108.L140501},
	journal = {Physical Review B},
	author = {Valentinis, Davide and Inkof, Gian Andrea and Schmalian, Jörg},
	year = {2023},
	pages = {L140501},
	note = {arXiv:2302.13138},
}

@article{valentinis_correlation_2023,
	title = {Correlation between phase stiffness and condensation energy across the non-{Fermi} to {Fermi}-liquid crossover in the {Yukawa}-{Sachdev}-{Ye}-{Kitaev} model on a lattice},
	volume = {5},
	url = {https://link.aps.org/doi/10.1103/PhysRevResearch.5.043007},
	doi = {10.1103/PhysRevResearch.5.043007},
	number = {4},
	urldate = {2026-07-17},
	journal = {Physical Review Research},
	publisher = {American Physical Society},
	author = {Valentinis, D. and Inkof, G. A. and Schmalian, J.},
	month = oct,
	year = {2023},
	pages = {043007},
}

@article{davis_quantum_2023,
	title = {Quantum chaos and phase transition in the {Yukawa}-{SYK} model},
	volume = {107},
	issn = {2469-9950, 2469-9969},
	url = {http://arxiv.org/abs/2212.03265},
	doi = {10.1103/PhysRevB.107.205122},
	number = {20},
	urldate = {2026-07-17},
	journal = {Physical Review B},
	author = {Davis, Andrew and Wang, Yuxuan},
	month = may,
	year = {2023},
	note = {arXiv:2212.03265 [cond-mat.str-el]},
	pages = {205122},
}

@article{li_strange_2024,
	title = {Strange metal and superconductor in the two-dimensional {Yukawa}-{Sachdev}-{Ye}-{Kitaev} model},
	volume = {133},
	issn = {0031-9007, 1079-7114},
	url = {http://arxiv.org/abs/2406.07608},
	doi = {10.1103/PhysRevLett.133.186502},
	number = {18},
	urldate = {2026-07-17},
	journal = {Physical Review Letters},
	author = {Li, Chenyuan and Valentinis, Davide and Patel, Aavishkar A. and Guo, Haoyu and Schmalian, Jörg and Sachdev, Subir and Esterlis, Ilya},
	month = oct,
	year = {2024},
	note = {arXiv:2406.07608 [cond-mat.str-el]},
	pages = {186502},
}

@article{bashan_tunable_2024,
	title = {Tunable non-{Fermi} liquid phase from coupling to two-level systems},
	volume = {132},
	issn = {0031-9007, 1079-7114},
	url = {http://arxiv.org/abs/2310.07768},
	doi = {10.1103/PhysRevLett.132.236501},
	number = {23},
	urldate = {2026-07-17},
	journal = {Physical Review Letters},
	author = {Bashan, Noga and Tulipman, Evyatar and Schmalian, Jörg and Berg, Erez},
	month = jun,
	year = {2024},
	note = {arXiv:2310.07768 [cond-mat.str-el]},
	pages = {236501},
}

@article{wang_quantum_2020,
	title = {Quantum {Phase} {Transition} in the {Yukawa}-{SYK} {Model}},
	volume = {2},
	issn = {2643-1564},
	url = {http://arxiv.org/abs/2005.07205},
	doi = {10.1103/PhysRevResearch.2.033084},
	number = {3},
	urldate = {2026-07-17},
	journal = {Physical Review Research},
	author = {Wang, Yuxuan and Chubukov, Andrey V.},
	month = jul,
	year = {2020},
	note = {arXiv:2005.07205 [cond-mat.str-el]},
	pages = {033084},
}

@article{chowdhury_sachdev-ye-kitaev_2022,
	title = {Sachdev-{Ye}-{Kitaev} {Models} and {Beyond}: {A} {Window} into {Non}-{Fermi} {Liquids}},
	volume = {94},
	issn = {0034-6861, 1539-0756},
	shorttitle = {Sachdev-{Ye}-{Kitaev} {Models} and {Beyond}},
	url = {http://arxiv.org/abs/2109.05037},
	doi = {10.1103/RevModPhys.94.035004},
	number = {3},
	urldate = {2026-07-17},
	journal = {Reviews of Modern Physics},
	author = {Chowdhury, Debanjan and Georges, Antoine and Parcollet, Olivier and Sachdev, Subir},
	month = sep,
	year = {2022},
	note = {arXiv:2109.05037 [cond-mat.str-el]},
	pages = {035004},
}

@article{pan_yukawa-syk_2021,
	title = {Yukawa-{SYK} model and {Self}-tuned {Quantum} {Criticality}},
	volume = {3},
	issn = {2643-1564},
	url = {http://arxiv.org/abs/2001.06586},
	doi = {10.1103/PhysRevResearch.3.013250},
	number = {1},
	urldate = {2026-07-17},
	journal = {Physical Review Research},
	author = {Pan, Gaopei and Wang, Wei and Davis, Andrew and Wang, Yuxuan and Meng, Zi Yang},
	month = mar,
	year = {2021},
	note = {arXiv:2001.06586 [cond-mat.str-el]},
	pages = {013250},
}

@article{chowdhury_intrinsic_2020,
	title = {Intrinsic superconducting instabilities of a solvable model for an incoherent metal},
	volume = {2},
	issn = {2643-1564},
	url = {http://arxiv.org/abs/1908.02757},
	doi = {10.1103/PhysRevResearch.2.013301},
	number = {1},
	urldate = {2026-07-17},
	journal = {Physical Review Research},
	author = {Chowdhury, Debanjan and Berg, Erez},
	month = mar,
	year = {2020},
	note = {arXiv:1908.02757 [cond-mat.str-el]},
	pages = {013301},
}

@article{esterlis_cooper_2019,
	title = {Cooper pairing of incoherent electrons: an electron-phonon version of the {Sachdev}-{Ye}-{Kitaev} model},
	volume = {100},
	issn = {2469-9950, 2469-9969},
	shorttitle = {Cooper pairing of incoherent electrons},
	url = {http://arxiv.org/abs/1906.04747},
	doi = {10.1103/PhysRevB.100.115132},
	number = {11},
	urldate = {2026-07-17},
	journal = {Physical Review B},
	author = {Esterlis, Ilya and Schmalian, Jörg},
	month = sep,
	year = {2019},
	note = {arXiv:1906.04747 [cond-mat.str-el]},
	pages = {115132},
}

@article{wang_solvable_2020,
	title = {Solvable {Strong}-{Coupling} {Quantum}-{Dot} {Model} with a {Non}-{Fermi}-{Liquid} {Pairing} {Transition}},
	volume = {124},
	doi = {10.1103/PhysRevLett.124.017002},
	number = {1},
	journal = {Physical Review Letters},
	author = {Wang, Yuxuan},
	year = {2020},
    pages = {017002},
}

@article{hauck_eliashberg_2020,
	title = {Eliashberg equations for an electron-phonon version of the {Sachdev}-{Ye}-{Kitaev} model: {Pair} {Breaking} in non-{Fermi} liquid superconductors},
	volume = {417},
	issn = {00034916},
	shorttitle = {Eliashberg equations for an electron-phonon version of the {Sachdev}-{Ye}-{Kitaev} model},
	url = {http://arxiv.org/abs/1911.04328},
	doi = {10.1016/j.aop.2020.168120},
	urldate = {2026-07-17},
	journal = {Annals of Physics},
	author = {Hauck, Daniel and Klug, Markus J. and Esterlis, Ilya and Schmalian, Jörg},
	month = jun,
	year = {2020},
	note = {arXiv:1911.04328 [cond-mat.str-el]},
	pages = {168120},
}

@article{wang_phase_2021,
	title = {Phase diagram of the spin-{$1/2$} {Yukawa}--{Sachdev}-{Ye}-{Kitaev} model: {Non}-{Fermi} liquid, insulator, and superconductor},
	volume = {103},
	doi = {10.1103/PhysRevB.103.195108},
	number = {19},
	journal = {Physical Review B},
	author = {Wang, Wei and Davis, Andrew and Pan, Gaopei and Wang, Yuxuan and Meng, Zi Yang},
	year = {2021},
	pages = {195108},
}

@article{classen_superconductivity_2021,
	title = {Superconductivity of incoherent electrons in {Yukawa}-{SYK} model},
	volume = {104},
	issn = {2469-9950, 2469-9969},
	url = {http://arxiv.org/abs/2106.12078},
	doi = {10.1103/PhysRevB.104.125120},
	number = {12},
	urldate = {2026-07-17},
	journal = {Physical Review B},
	author = {Classen, Laura and Chubukov, Andrey},
	month = sep,
	year = {2021},
	note = {arXiv:2106.12078 [cond-mat.str-el]},
	pages = {125120},
}

@article{bi_instability_2017,
	title = {Instability of the non-{Fermi} liquid state of the {Sachdev}-{Ye}-{Kitaev} {Model}},
	volume = {95},
	issn = {2469-9950, 2469-9969},
	url = {http://arxiv.org/abs/1701.07081},
	doi = {10.1103/PhysRevB.95.205105},
	number = {20},
	urldate = {2026-07-17},
	journal = {Physical Review B},
	author = {Bi, Zhen and Jian, Chao-Ming and You, Yi-Zhuang and Pawlak, Kelly Ann and Xu, Cenke},
	month = may,
	year = {2017},
	note = {arXiv:1701.07081 [cond-mat.str-el]},
	pages = {205105},
}

@article{chubukov_superconductivity_2005,
	title = {Superconductivity due to massless boson exchange in the strong-coupling limit},
	volume = {72},
	url = {https://link.aps.org/doi/10.1103/PhysRevB.72.174520},
	doi = {10.1103/PhysRevB.72.174520},
	number = {17},
	urldate = {2026-07-17},
	journal = {Physical Review B},
	publisher = {American Physical Society},
	author = {Chubukov, Andrey V. and Schmalian, Jörg},
	month = nov,
	year = {2005},
	pages = {174520},
}

@article{klebanov_uncolored_2017,
	title = {Uncolored {Random} {Tensors}, {Melon} {Diagrams}, and the {SYK} {Models}},
	volume = {95},
	issn = {2470-0010, 2470-0029},
	url = {http://arxiv.org/abs/1611.08915},
	doi = {10.1103/PhysRevD.95.046004},
	number = {4},
	urldate = {2026-07-18},
	journal = {Physical Review D},
	author = {Klebanov, Igor R. and Tarnopolsky, Grigory},
	month = feb,
	year = {2017},
	note = {arXiv:1611.08915 [hep-th]},
	pages = {046004},
}

@article{peng_vector_2017,
	title = {Vector models and generalized {SYK} models},
	volume = {2017},
	issn = {1029-8479},
	url = {http://arxiv.org/abs/1704.04223},
	doi = {10.1007/JHEP05(2017)129},
	number = {5},
	urldate = {2026-07-18},
	journal = {Journal of High Energy Physics},
	author = {Peng, Cheng},
	month = may,
	year = {2017},
	note = {arXiv:1704.04223 [hep-th]},
	pages = {129},
}

@misc{fraser-taliente_melonic_2024,
	title = {Melonic limits of the quartic {Yukawa} model and general features of melonic {CFTs}},
	url = {http://arxiv.org/abs/2410.09152},
	doi = {10.48550/arXiv.2410.09152},
	urldate = {2026-07-18},
	publisher = {arXiv},
	author = {Fraser-Taliente, Ludo and Wheater, John},
	month = oct,
	year = {2024},
	note = {arXiv:2410.09152 [hep-th]},
}

@article{migdal_interaction_1958,
	title = {Interaction between electrons and lattice vibrations in a normal metal},
	volume = {7},
	journal = {Soviet Physics JETP},
	author = {Migdal, A. B.},
	year = {1958},
	pages = {996},
	note = {[Zh. Eksp. Teor. Fiz. 34, 1438 (1958)]},
}

@article{eliashberg_interactions_1960,
	title = {Interactions between electrons and lattice vibrations in a superconductor},
	volume = {11},
	journal = {Soviet Physics JETP},
	author = {Eliashberg, G. M.},
	year = {1960},
	pages = {696},
	note = {[Zh. Eksp. Teor. Fiz. 38, 966 (1960)]},
}

@article{eliashberg_temperature_1963,
	title = {TITLE-TO-FILL},
	volume = {16},
	journal = {Soviet Physics JETP},
	author = {Eliashberg, G. M.},
	year = {1963},
	pages = {780},
	note = {[Zh. Eksp. Teor. Fiz. 43, 1005 (1962)]},
}

@article{alexandrov_breakdown_2001,
	title = {Breakdown of the {Migdal}-{Eliashberg} theory in the strong-coupling adiabatic regime},
	volume = {56},
	issn = {0295-5075, 1286-4854},
	url = {http://arxiv.org/abs/cond-mat/0102189},
	doi = {10.1209/epl/i2001-00492-x},
	number = {1},
	urldate = {2026-07-18},
	journal = {Europhysics Letters (EPL)},
	author = {Alexandrov, A. S.},
	month = oct,
	year = {2001},
	note = {arXiv:cond-mat/0102189},
	pages = {92--98},
}

@article{marsiglio_eliashberg_2020,
	title = {Eliashberg {Theory}: a short review},
	volume = {417},
	issn = {00034916},
	shorttitle = {Eliashberg {Theory}},
	url = {http://arxiv.org/abs/1911.05065},
	doi = {10.1016/j.aop.2020.168102},
	urldate = {2026-07-18},
	journal = {Annals of Physics},
	author = {Marsiglio, F.},
	month = jun,
	year = {2020},
	note = {arXiv:1911.05065 [cond-mat.supr-con]},
	pages = {168102},
}

@article{bauer_quantitative_2011,
	title = {Quantitative reliability of {Migdal}-{Eliashberg} theory for strong electron-phonon coupling},
	volume = {84},
	issn = {1098-0121, 1550-235X},
	url = {http://arxiv.org/abs/1105.2833},
	doi = {10.1103/PhysRevB.84.184531},
	number = {18},
	urldate = {2026-07-18},
	journal = {Physical Review B},
	author = {Bauer, Johannes and Han, Jong E. and Gunnarsson, Olle},
	month = nov,
	year = {2011},
	note = {arXiv:1105.2833 [cond-mat.supr-con]},
	pages = {184531},
}

@article{chowdhury_unreasonable_2020,
	title = {The unreasonable effectiveness of {Eliashberg} theory for pairing of non-{Fermi} liquids},
	volume = {417},
	issn = {00034916},
	url = {http://arxiv.org/abs/1912.07646},
	doi = {10.1016/j.aop.2020.168125},
	urldate = {2026-07-18},
	journal = {Annals of Physics},
	author = {Chowdhury, Debanjan and Berg, Erez},
	month = jun,
	year = {2020},
	note = {arXiv:1912.07646 [cond-mat.supr-con]},
	pages = {168125},
}

@article{yang_non-perturbative_2023,
	title = {A non-perturbative study of the interplay between electron-phonon interaction and {Coulomb} interaction in undoped graphene},
	volume = {35},
	issn = {0953-8984, 1361-648X},
	url = {http://arxiv.org/abs/2201.01472},
	doi = {10.1088/1361-648X/aca948},
	number = {7},
	urldate = {2026-07-18},
	journal = {Journal of Physics: Condensed Matter},
	author = {Yang, Zhao-Kun and Pan, Xiao-Yin and Liu, Guo-Zhu},
	month = feb,
	year = {2023},
	note = {arXiv:2201.01472 [cond-mat.str-el]},
	pages = {075601},
}

@article{pan_nonperturbative_2021,
	title = {Nonperturbative {Dyson}-{Schwinger} equation approach to strongly interacting {Dirac} fermion systems},
	volume = {104},
	issn = {2469-9950, 2469-9969},
	url = {http://arxiv.org/abs/2003.10371},
	doi = {10.1103/PhysRevB.104.085141},
	number = {8},
	urldate = {2026-07-18},
	journal = {Physical Review B},
	author = {Pan, Xiao-Yin and Yang, Zhao-Kun and Li, Xin and Liu, Guo-Zhu},
	month = aug,
	year = {2021},
	note = {arXiv:2003.10371 [cond-mat.str-el]},
	pages = {085141},
}

@article{huang_theoretical_2023,
	title = {Theoretical study of phonon-mediated superconductivity beyond {Migdal}-{Eliashberg} approximation and {Coulomb} pseudopotential},
	volume = {108},
	issn = {2469-9950, 2469-9969},
	url = {http://arxiv.org/abs/2301.13520},
	doi = {10.1103/PhysRevB.108.144507},
	number = {14},
	urldate = {2026-07-18},
	journal = {Physical Review B},
	author = {Huang, Jie and Yang, Zhao-Kun and Pan, Xiao-Yin and Liu, Guo-Zhu},
	month = oct,
	year = {2023},
	note = {arXiv:2301.13520 [cond-mat.supr-con]},
	pages = {144507},
}

@article{huang_superconductivity_2025,
	title = {Superconductivity near an {Ising} nematic quantum critical point in two dimensions},
	volume = {112},
	issn = {2469-9950, 2469-9969},
	url = {http://arxiv.org/abs/2502.08270},
	doi = {10.1103/6n9b-wzyk},
	number = {21},
	urldate = {2026-07-18},
	journal = {Physical Review B},
	author = {Huang, Jie and Yang, Zhao-Kun and Wang, Jing-Rong and Liu, Guo-Zhu},
	month = dec,
	year = {2025},
	note = {arXiv:2502.08270 [cond-mat.supr-con]},
	pages = {214516},
}

@article{schrodi_full-bandwidth_2020,
	title = {Full-bandwidth {Eliashberg} theory of superconductivity beyond {Migdal}'s approximation},
	volume = {102},
	issn = {2469-9950, 2469-9969},
	url = {http://arxiv.org/abs/1911.12872},
	doi = {10.1103/PhysRevB.102.024503},
	number = {2},
	urldate = {2026-07-18},
	journal = {Physical Review B},
	author = {Schrodi, Fabian and Oppeneer, Peter M. and Aperis, Alex},
	month = jul,
	year = {2020},
	note = {arXiv:1911.12872 [cond-mat.supr-con]},
	pages = {024503},
}

@article{aldape_solvable_2022,
	title = {Solvable {Theory} of a {Strange} {Metal} at the {Breakdown} of a {Heavy} {Fermi} {Liquid}},
	volume = {105},
	issn = {2469-9950, 2469-9969},
	url = {http://arxiv.org/abs/2012.00763},
	doi = {10.1103/PhysRevB.105.235111},
	number = {23},
	urldate = {2026-07-18},
	journal = {Physical Review B},
	author = {Aldape, Erik E. and Cookmeyer, Tessa and Patel, Aavishkar A. and Altman, Ehud},
	month = jun,
	year = {2022},
	note = {arXiv:2012.00763 [cond-mat.str-el]},
	pages = {235111},
}

@article{damia_thermal_2021,
	title = {Thermal effects in non-{Fermi} liquid superconductivity},
	volume = {103},
	issn = {2469-9950, 2469-9969},
	url = {http://arxiv.org/abs/2009.11887},
	doi = {10.1103/PhysRevB.103.155161},
	number = {15},
	urldate = {2026-07-18},
	journal = {Physical Review B},
	author = {Damia, Jeremias Aguilera and Solis, Mario and Torroba, Gonzalo},
	month = apr,
	year = {2021},
	note = {arXiv:2009.11887 [cond-mat.str-el]},
	pages = {155161},
}

@article{wang_non-fermi_2017,
	title = {Non-{Fermi} liquid {Superconductivity}: {Eliashberg} versus the {Renormalization} {Group}},
	volume = {95},
	issn = {2469-9950, 2469-9969},
	shorttitle = {Non-{Fermi} liquid {Superconductivity}},
	url = {http://arxiv.org/abs/1612.01971},
	doi = {10.1103/PhysRevB.95.165137},
	number = {16},
	urldate = {2026-07-18},
	journal = {Physical Review B},
	author = {Wang, Huajia and Raghu, Srinivas and Torroba, Gonzalo},
	month = apr,
	year = {2017},
	note = {arXiv:1612.01971 [cond-mat.str-el]},
	pages = {165137},
}

@article{wang_superconductivity_2018,
	title = {Superconductivity vs quantum criticality: effects of thermal fluctuations},
	volume = {97},
	issn = {2469-9950, 2469-9969},
	shorttitle = {Superconductivity vs quantum criticality},
	url = {http://arxiv.org/abs/1708.04624},
	doi = {10.1103/PhysRevB.97.054502},
	number = {5},
	urldate = {2026-07-18},
	journal = {Physical Review B},
	author = {Wang, Huajia and Wang, Yuxuan and Torroba, Gonzalo},
	month = feb,
	year = {2018},
	note = {arXiv:1708.04624 [cond-mat.supr-con]},
	pages = {054502},
}

@misc{witten_syk-like_2016,
	title = {An {SYK}-{Like} {Model} {Without} {Disorder}},
	url = {http://arxiv.org/abs/1610.09758},
	doi = {10.48550/arXiv.1610.09758},
	urldate = {2026-07-18},
	publisher = {arXiv},
	author = {Witten, Edward},
	month = nov,
	year = {2016},
	note = {arXiv:1610.09758 [hep-th]},
}

@article{gurau_complete_2017,
	title = {The complete \$1/{N}\$ expansion of a {SYK}--like tensor model},
	volume = {916},
	issn = {05503213},
	url = {http://arxiv.org/abs/1611.04032},
	doi = {10.1016/j.nuclphysb.2017.01.015},
	urldate = {2026-07-18},
	journal = {Nuclear Physics B},
	author = {Gurau, Razvan},
	month = mar,
	year = {2017},
	note = {arXiv:1611.04032 [hep-th]},
	pages = {386--401},
}

@article{gurau_1n_2011,
	title = {The 1/{N} expansion of colored tensor models},
	volume = {12},
	issn = {1424-0637, 1424-0661},
	url = {http://arxiv.org/abs/1011.2726},
	doi = {10.1007/s00023-011-0101-8},
	number = {5},
	urldate = {2026-07-18},
	journal = {Annales Henri Poincaré},
	author = {Gurau, Razvan},
	month = jul,
	year = {2011},
	note = {arXiv:1011.2726 [gr-qc]},
	pages = {829--847},
}

@article{tsoncheva_condensation_2005,
	title = {Condensation energy in {Eliashberg} theory -- from weak to strong coupling},
	volume = {71},
	issn = {1098-0121, 1550-235X},
	url = {http://arxiv.org/abs/cond-mat/0503512},
	doi = {10.1103/PhysRevB.71.184513},
	number = {18},
	urldate = {2026-07-18},
	journal = {Physical Review B},
	author = {Tsoncheva, Evelina and Chubukov, Andrey V.},
	month = may,
	year = {2005},
	note = {arXiv:cond-mat/0503512},
	pages = {184513},
}

@misc{marsiglio_electron_2001,
	title = {Electron - {Phonon} {Superconductivity}},
	url = {http://arxiv.org/abs/cond-mat/0106143},
	doi = {10.48550/arXiv.cond-mat/0106143},
	urldate = {2026-07-18},
	publisher = {arXiv},
	author = {Marsiglio, F. and Carbotte, J. P.},
	month = jun,
	year = {2001},
	note = {arXiv:cond-mat/0106143},
}

@article{mirabi_thermodynamics_2020,
	title = {Thermodynamics of {Eliashberg} theory in the weak-coupling limit},
	volume = {102},
	issn = {2469-9950, 2469-9969},
	url = {http://arxiv.org/abs/2010.02801},
	doi = {10.1103/PhysRevB.102.214505},
	number = {21},
	urldate = {2026-07-18},
	journal = {Physical Review B},
	author = {Mirabi, Sepideh and Boyack, Rufus and Marsiglio, F.},
	month = dec,
	year = {2020},
	note = {arXiv:2010.02801 [cond-mat.supr-con]},
	pages = {214505},
}

@article{zhang_superconductivity_2022,
	title = {Superconductivity out of a non-{Fermi} liquid. {Free} energy analysis},
	volume = {106},
	issn = {2469-9950, 2469-9969},
	url = {http://arxiv.org/abs/2208.13888},
	doi = {10.1103/PhysRevB.106.144513},
	number = {14},
	urldate = {2026-07-18},
	journal = {Physical Review B},
	author = {Zhang, Shang-Shun and Wu, Yi-Ming and Abanov, Artem and Chubukov, Andrey V.},
	month = oct,
	year = {2022},
	note = {arXiv:2208.13888 [cond-mat.supr-con]},
	pages = {144513},
}

@article{zhang_free_2023-1,
	title = {Free energy and specific heat near a quantum critical point of a metal},
	volume = {107},
	issn = {2469-9950, 2469-9969},
	url = {http://arxiv.org/abs/2301.01873},
	doi = {10.1103/PhysRevB.107.144507},
	number = {14},
	urldate = {2026-07-18},
	journal = {Physical Review B},
	author = {Zhang, Shang-Shun and Berg, Erez and Chubukov, Andrey V.},
	month = apr,
	year = {2023},
	note = {arXiv:2301.01873 [cond-mat.str-el]},
	pages = {144507},
}

@article{paci_polaronic_2005,
	title = {Polaronic and nonadiabatic phase diagram from anomalous isotope effects},
	volume = {94},
	issn = {0031-9007},
	doi = {10.1103/PhysRevLett.94.036406},
	number = {3},
	journal = {Physical Review Letters},
	author = {Paci, P. and Capone, M. and Cappelluti, E. and Ciuchi, S. and Grimaldi, C. and Pietronero, L.},
	month = jan,
	year = {2005},
	pages = {036406},
}

@article{fratini_optical_2006,
	title = {Optical properties of small polarons from dynamical mean-field theory},
	volume = {74},
	url = {https://link.aps.org/doi/10.1103/PhysRevB.74.075101},
	doi = {10.1103/PhysRevB.74.075101},
	number = {7},
	urldate = {2026-07-18},
	journal = {Physical Review B},
	publisher = {American Physical Society},
	author = {Fratini, S. and Ciuchi, S.},
	month = aug,
	year = {2006},
	pages = {075101},
}

@article{heidarian_inhomogeneous_2003,
	title = {Inhomogeneous metallic phase in a disordered {Mott} insulator in two dimensions},
	volume = {93},
	doi = {10.1103/PhysRevLett.93.126401},
	number = {12},
	journal = {Physical Review Letters},
	author = {Heidarian, Dariush and Trivedi, Nandini},
	year = {2004},
	pages = {126401},
}

@misc{wolfle_self-consistent_2010,
	title = {Self-{Consistent} {Theory} of {Anderson} {Localization}: {General} {Formalism} and {Applications}},
	shorttitle = {Self-{Consistent} {Theory} of {Anderson} {Localization}},
	url = {https://arxiv.org/abs/1004.3238v2},
	doi = {10.1142/S0217979210064502},
	urldate = {2026-07-18},
	journal = {arXiv.org},
	author = {Wölfle, P. and Vollhardt, D.},
	month = apr,
	year = {2010},
}

@misc{miskic_polaron_2025,
	title = {Polaron formation as the vertex function problem: {From} {Dyck}'s paths to self-energy {Feynman} diagrams},
	shorttitle = {Polaron formation as the vertex function problem},
	url = {https://arxiv.org/abs/2505.21054v3},
	doi = {10.21468/SciPostPhys.20.5.129},
	urldate = {2026-07-18},
	journal = {arXiv.org},
	author = {Miškić, Tomislav and Krsnik, Juraj and Ragni, Stefano and Mishchenko, Andrey S. and Barišić, Osor S.},
	month = may,
	year = {2025},
}

@misc{ragni_polarons_2025,
	title = {Polarons with arbitrary nonlinear electron-phonon interaction},
	url = {https://arxiv.org/abs/2506.17914v1},
	doi = {10.1103/6127-phps},
	urldate = {2026-07-18},
	journal = {arXiv.org},
	author = {Ragni, Stefano and Miškić, Tomislav and Hahn, Thomas and Prokof'ev, Nikolay and Barišić, Osor S. and Nagaosa, Naoto and Franchini, Cesare and Mishchenko, Andrey S.},
	month = jun,
	year = {2025},
}

@misc{chubukov_breakdown_2026,
	title = {Breakdown of the {Migdal}-{Eliashberg} theory for electron-phonon systems. {Role} of polarons/bi-polarons},
	url = {https://arxiv.org/abs/2604.14294v2},
	urldate = {2026-07-18},
	journal = {arXiv.org},
	author = {Chubukov, Andrey and Esterlis, Ilya and Abanov, Artem and Prokof'ev, Nikolay},
	month = apr,
	year = {2026},
}

@misc{prokofev_limits_2026,
	title = {Limits of validity for {Migdal}-{Eliashberg} theory: role of polarons/bi-polarons},
	shorttitle = {Limits of validity for {Migdal}-{Eliashberg} theory},
	doi = {10.48550/arXiv.2604.14293},
	author = {Prokof'ev, Nikolay and Esterlis, Ilya and Abanov, Ar and Chubukov, Andrey},
	month = apr,
	year = {2026},
}

@article{chubukov_eliashberg_2020,
	title = {Eliashberg theory of phonon-mediated superconductivity -- when it is valid and how it breaks down},
	volume = {417},
	doi = {10.1016/j.aop.2020.168190},
	journal = {Annals of Physics},
	author = {Chubukov, Andrey V. and Abanov, Artem and Esterlis, Ilya and Kivelson, Steven A.},
	year = {2020},
	pages = {168190},
}

@misc{cai_superconductivity_2025,
	title = {Superconductivity in {Electron} {Liquids}: {Precision} {Many}-{Body} {Treatment} of {Coulomb} {Interaction}},
	copyright = {arXiv.org perpetual, non-exclusive license},
	shorttitle = {Superconductivity in {Electron} {Liquids}},
	url = {https://arxiv.org/abs/2512.19382},
	doi = {10.48550/ARXIV.2512.19382},
	urldate = {2026-07-18},
	publisher = {arXiv},
	author = {Cai, Xiansheng and Wang, Tao and Zhang, Shuai and Zhang, Tiantian and Millis, Andrew and Svistunov, Boris V. and Prokof'ev, Nikolay V. and Chen, Kun},
	year = {2025},
	note = {Version Number: 2},
}

@misc{kabanov_adiabatic_2020,
	title = {Adiabatic theory of the polaron spectral function},
	url = {https://arxiv.org/abs/2011.08638v3},
	doi = {10.1088/2399-6528/ac9d81},
	urldate = {2026-07-18},
	journal = {arXiv.org},
	author = {Kabanov, V. V.},
	month = nov,
	year = {2020},
}

@article{yuzbashyan_superconductivity_2022,
	title = {Superconductivity near a quantum critical point in the extreme retardation regime},
	volume = {106},
	issn = {2469-9950, 2469-9969},
	url = {http://arxiv.org/abs/2206.07575},
	doi = {10.1103/PhysRevB.106.064502},
	number = {6},
	urldate = {2026-07-18},
	journal = {Physical Review B},
	author = {Yuzbashyan, Emil A. and Kiessling, Michael K.-H. and Altshuler, Boris L.},
	month = aug,
	year = {2022},
	note = {arXiv:2206.07575 [cond-mat.supr-con]},
	pages = {064502},
}

@article{yuzbashyan_breakdown_2022,
	title = {Breakdown of the {Migdal}-{Eliashberg} theory and a theory of lattice-fermionic superfluidity},
	volume = {106},
	issn = {2469-9950, 2469-9969},
	url = {http://arxiv.org/abs/2206.01593},
	doi = {10.1103/PhysRevB.106.054518},
	number = {5},
	urldate = {2026-07-18},
	journal = {Physical Review B},
	author = {Yuzbashyan, Emil A. and Altshuler, Boris L.},
	month = aug,
	year = {2022},
	note = {arXiv:2206.01593 [cond-mat.supr-con]},
	pages = {054518},
}

@article{yuzbashyan_migdal-eliashberg_2022,
	title = {Migdal-{Eliashberg} theory as a classical spin chain},
	volume = {106},
	issn = {2469-9950, 2469-9969},
	url = {http://arxiv.org/abs/2205.06442},
	doi = {10.1103/PhysRevB.106.014512},
	number = {1},
	urldate = {2026-07-18},
	journal = {Physical Review B},
	author = {Yuzbashyan, Emil A. and Altshuler, Boris L.},
	month = jul,
	year = {2022},
	note = {arXiv:2205.06442 [cond-mat.supr-con]},
	pages = {014512},
}

@article{yuzbashyan_instability_2025,
	title = {Instability of metals with respect to strong electron-phonon interaction},
	volume = {135},
	issn = {0031-9007, 1079-7114},
	url = {http://arxiv.org/abs/2409.19562},
	doi = {10.1103/sf5p-2g5l},
	number = {2},
	urldate = {2026-07-18},
	journal = {Physical Review Letters},
	author = {Yuzbashyan, Emil A. and Altshuler, Boris L. and Patra, Aniket},
	month = jul,
	year = {2025},
	note = {arXiv:2409.19562 [cond-mat.str-el]},
	pages = {026503},
}

@article{kiessling_bounds_2025,
	title = {Bounds on {$T_c$} in the {Eliashberg} theory of {Superconductivity}. {I}: {The} {$\gamma$}-model},
	issn = {1572-9613},
	shorttitle = {Bounds on \${T}\_c\$ in the {Eliashberg} theory of {Superconductivity}. {I}},
	url = {http://arxiv.org/abs/2409.00533},
	doi = {10.1007/s10955-025-03446-5},
	number = {5},
	urldate = {2026-07-18},
	journal = {Journal of Statistical Physics},
	author = {Kiessling, Michael K.-H. and Altshuler, Boris L. and Yuzbashyan, Emil A.},
	month = may,
	year = {2025},
	note = {arXiv:2409.00533 [math-ph]},
	pages = {69},
}

@article{kiessling_bounds_2025-1,
	title = {Bounds on {$T_c$} in the {Eliashberg} theory of {Superconductivity}. {II}: {Dispersive} phonons},
	volume = {192},
	issn = {1572-9613},
	shorttitle = {Bounds on \${T}\_c\$ in the {Eliashberg} theory of {Superconductivity}. {II}},
	url = {http://arxiv.org/abs/2409.00532},
	doi = {10.1007/s10955-025-03468-z},
	number = {7},
	urldate = {2026-07-18},
	journal = {Journal of Statistical Physics},
	author = {Kiessling, Michael K.-H. and Altshuler, Boris L. and Yuzbashyan, Emil A.},
	month = jul,
	year = {2025},
	note = {arXiv:2409.00532 [math-ph]},
	pages = {94},
}

@article{kiessling_bounds_2025-2,
	title = {Bounds on {$T_c$} in the {Eliashberg} theory of {Superconductivity}. {III}: {Einstein} phonons},
	issn = {1572-9613},
	shorttitle = {Bounds on \${T}\_c\$ in the {Eliashberg} theory of {Superconductivity}. {III}},
	url = {http://arxiv.org/abs/2409.02121},
	doi = {10.1007/s10955-025-03469-y},
	number = {7},
	urldate = {2026-07-18},
	journal = {Journal of Statistical Physics},
	author = {Kiessling, Michael K.-H. and Altshuler, Boris L. and Yuzbashyan, Emil A.},
	month = jul,
	year = {2025},
	note = {arXiv:2409.02121 [cond-mat.supr-con]},
	pages = {93},
}

@article{semenok_fundamental_2025,
	title = {Fundamental limits on the electron-phonon coupling and superconducting {$T_c$}},
	volume = {37},
	issn = {0935-9648, 1521-4095},
	url = {http://arxiv.org/abs/2407.12922},
	doi = {10.1002/adma.202507013},
	number = {40},
	urldate = {2026-07-18},
	journal = {Advanced Materials},
	author = {Semenok, Dmitrii V. and Altshuler, Boris L. and Yuzbashyan, Emil A.},
	month = oct,
	year = {2025},
	note = {arXiv:2407.12922 [cond-mat.supr-con]},
	pages = {2507013},
}

@article{anderson_new_1958,
	title = {New Method in the Theory of Superconductivity},
	volume = {110},
	doi = {10.1103/PhysRev.110.985},
	number = {4},
	journal = {Physical Review},
	author = {Anderson, P. W.},
	month = may,
	year = {1958},
	pages = {985--986},
}

@article{anderson_random-phase_1958,
	title = {Random-Phase Approximation in the Theory of Superconductivity},
	volume = {112},
	doi = {10.1103/PhysRev.112.1900},
	number = {6},
	journal = {Physical Review},
	author = {Anderson, P. W.},
	month = dec,
	year = {1958},
	pages = {1900--1916},
}

@misc{khveshchenko_unpublished,
	author = {Khveshchenko, D. V.},
	note = {unpublished},
    year = {2026},
}

@article{appelquist_spontaneous_1986,
	title = {Spontaneous chiral-symmetry breaking in three-dimensional {QED}},
	volume = {33},
	url = {https://link.aps.org/doi/10.1103/PhysRevD.33.3704},
	doi = {10.1103/PhysRevD.33.3704},
	number = {12},
	urldate = {2026-07-19},
	journal = {Physical Review D},
	publisher = {American Physical Society},
	author = {Appelquist, Thomas W. and Bowick, Mark and Karabali, Dimitra and Wijewardhana, L. C. R.},
	month = jun,
	year = {1986},
	pages = {3704--3713},
}

@article{leal_excitonic_2004,
	title = {Excitonic instability in two-dimensional degenerate semimetals},
	volume = {687},
	issn = {05503213},
	url = {http://arxiv.org/abs/cond-mat/0302164},
	doi = {10.1016/j.nuclphysb.2004.03.020},
	number = {3},
	urldate = {2026-07-19},
	journal = {Nuclear Physics B},
	author = {Leal, H. and Khveshchenko, D. V.},
	month = may,
	year = {2004},
	note = {arXiv:cond-mat/0302164},
	pages = {323--331},
}

@article{khveshchenko_ghost_2001,
	title = {Ghost {Excitonic} {Insulator} {Transition} in {Layered} {Graphite}},
	volume = {87},
	url = {https://link.aps.org/doi/10.1103/PhysRevLett.87.246802},
	doi = {10.1103/PhysRevLett.87.246802},
	number = {24},
	urldate = {2026-07-19},
	journal = {Physical Review Letters},
	publisher = {American Physical Society},
	author = {Khveshchenko, D. V.},
	month = nov,
	year = {2001},
	pages = {246802},
}

@article{gorbar_magnetic_2002,
	title = {Magnetic field driven metal-insulator phase transition in planar systems},
	volume = {66},
	url = {https://link.aps.org/doi/10.1103/PhysRevB.66.045108},
	doi = {10.1103/PhysRevB.66.045108},
	number = {4},
	urldate = {2026-07-19},
	journal = {Physical Review B},
	publisher = {American Physical Society},
	author = {Gorbar, E. V. and Gusynin, V. P. and Miransky, V. A. and Shovkovy, I. A.},
	month = jul,
	year = {2002},
	pages = {045108},
}

@article{khveshchenko_excitonic_2006,
	title = {Excitonic pairing between nodal fermions},
	volume = {73},
	issn = {1098-0121, 1550-235X},
	url = {http://arxiv.org/abs/cond-mat/0510519},
	doi = {10.1103/PhysRevB.73.115104},
	number = {11},
	urldate = {2026-07-19},
	journal = {Physical Review B},
	author = {Khveshchenko, D. V. and Shively, W. F.},
	month = mar,
	year = {2006},
	note = {arXiv:cond-mat/0510519},
	pages = {115104},
}

@article{khveshchenko_massive_2009,
	title = {Massive {Dirac} fermions in single-layer graphene},
	volume = {21},
	issn = {0953-8984},
	url = {https://doi.org/10.1088/0953-8984/21/7/075303},
	doi = {10.1088/0953-8984/21/7/075303},
	number = {7},
	urldate = {2026-07-19},
	journal = {Journal of Physics: Condensed Matter},
	author = {Khveshchenko, D V},
	month = jan,
	year = {2009},
	pages = {075303},
}

@article{abanov_interplay_2020,
	title = {Interplay between superconductivity and non-{Fermi} liquid at a quantum-critical point in a metal. {I}: {The} {$\gamma$}-model and its phase diagram at {$T=0$}. {The} case {$0<\gamma<1$}},
	volume = {102},
	issn = {2469-9950, 2469-9969},
	shorttitle = {Interplay between superconductivity and non-{Fermi} liquid at a quantum-critical point in a metal. {I}},
	url = {http://arxiv.org/abs/2004.13220},
	doi = {10.1103/PhysRevB.102.024524},
	number = {2},
	urldate = {2026-07-19},
	journal = {Physical Review B},
	author = {Abanov, Artem and Chubukov, Andrey V.},
	month = jul,
	year = {2020},
	note = {arXiv:2004.13220 [cond-mat.str-el]},
	pages = {024524},
}

@article{wu_interplay_2020,
	title = {Interplay between superconductivity and non-{Fermi} liquid at a quantum critical point in a metal. {II}. {The} {$\gamma$}-model at a finite {$T$} for {$0<\gamma<1$}},
	volume = {102},
	issn = {2469-9950, 2469-9969},
	url = {http://arxiv.org/abs/2006.02968},
	doi = {10.1103/PhysRevB.102.024525},
	number = {2},
	urldate = {2026-07-19},
	journal = {Physical Review B},
	author = {Wu, Yiming and Abanov, Artem and Wang, Yuxuan and Chubukov, Andrey V.},
	month = jul,
	year = {2020},
	note = {arXiv:2006.02968 [cond-mat.supr-con]},
	pages = {024525},
}

@article{wu_interplay_2020-1,
	title = {Interplay between superconductivity and non-{Fermi} liquid at a quantum-critical point in a metal. {III}: {The} {$\gamma$} model and its phase diagram across {$\gamma=1$}},
	issn = {2469-9950, 2469-9969},
	shorttitle = {Interplay between superconductivity and non-{Fermi} liquid at a quantum-critical point in a metal. {III}},
	url = {http://arxiv.org/abs/2007.14540},
	doi = {10.1103/PhysRevB.102.094516},
	number = {9},
	urldate = {2026-07-19},
	journal = {Physical Review B},
	author = {Wu, Yi-Ming and Abanov, Artem and Chubukov, Andrey V.},
	month = sep,
	year = {2020},
	note = {arXiv:2007.14540 [cond-mat.supr-con]},
	pages = {094516},
}

@article{wu_interplay_2021,
	title = {Interplay between superconductivity and non-{Fermi} liquid at a quantum-critical point in a metal. {IV}: {The} {$\gamma$} model and its phase diagram at {$1<\gamma<2$}},
	volume = {103},
	issn = {2469-9950, 2469-9969},
	shorttitle = {Interplay between superconductivity and non-{Fermi} liquid at a quantum-critical point in a metal. {IV}},
	url = {http://arxiv.org/abs/2009.10911},
	doi = {10.1103/PhysRevB.103.024522},
	number = {2},
	urldate = {2026-07-19},
	journal = {Physical Review B},
	author = {Wu, Yi-Ming and Zhang, Shang-Shun and Abanov, Artem and Chubukov, Andrey V.},
	month = jan,
	year = {2021},
	note = {arXiv:2009.10911 [cond-mat.supr-con]},
	pages = {024522},
}

@article{wu_interplay_2021-1,
	title = {Interplay between superconductivity and non-{Fermi} liquid at a quantum-critical point in a metal: {V}. {The} {$\gamma$} model and its phase diagram. {The} case {$\gamma=2$}},
	volume = {103},
	issn = {2469-9950, 2469-9969},
	shorttitle = {Interplay between superconductivity and non-{Fermi} liquid at a quantum-critical point in a metal},
	url = {http://arxiv.org/abs/2102.04444},
	doi = {10.1103/PhysRevB.103.184508},
	number = {18},
	urldate = {2026-07-19},
	journal = {Physical Review B},
	author = {Wu, Yi-Ming and Zhang, Shang-Shun and Abanov, Artem and Chubukov, Andrey V.},
	month = may,
	year = {2021},
	note = {arXiv:2102.04444 [cond-mat.supr-con]},
	pages = {184508},
}

@article{zhang_interplay_2021,
	title = {Interplay between superconductivity and non-{Fermi} liquid at a quantum-critical point in a metal. {VI}. {The} {$\gamma$} model and its phase diagram at {$2<\gamma<3$}},
	volume = {104},
	issn = {2469-9950, 2469-9969},
	url = {http://arxiv.org/abs/2107.14340},
	doi = {10.1103/PhysRevB.104.144509},
	number = {14},
	urldate = {2026-07-19},
	journal = {Physical Review B},
	author = {Zhang, Shang-Shun and Wu, Yi-Ming and Abanov, Artem and Chubukov, Andrey V.},
	month = oct,
	year = {2021},
	note = {arXiv:2107.14340 [cond-mat.supr-con]},
	pages = {144509},
}

@misc{yu_topologically_2026,
	title = {Topologically non-trivial gap function and topology-induced time-reversal symmetry breaking in a superconductor with singular dynamical interaction},
	url = {http://arxiv.org/abs/2604.14295},
	doi = {10.48550/arXiv.2604.14295},
	urldate = {2026-07-19},
	publisher = {arXiv},
	author = {Yu, Yue and Chubukov, Andrey V.},
	month = apr,
	year = {2026},
	note = {arXiv:2604.14295 [cond-mat.str-el]},
}

@misc{solis_single-particle_2025,
	title = {From single-particle to many-body chaos in {Yukawa}--{SYK}: theory and a cavity-{QED} proposal},
	shorttitle = {From single-particle to many-body chaos in {Yukawa}--{SYK}},
	url = {http://arxiv.org/abs/2511.04762},
	doi = {10.48550/arXiv.2511.04762},
	urldate = {2026-07-20},
	publisher = {arXiv},
	author = {Solis, David Pascual and Windey, Alex and Bandyopadhyay, Soumik and Legramandi, Andrea and Hauke, Philipp},
	month = nov,
	year = {2025},
	note = {arXiv:2511.04762 [quant-ph]},
}

@article{zhang_gauging_2025,
	title = {Gauging the complex {SYK} model},
	volume = {2025},
	issn = {1029-8479},
	url = {http://arxiv.org/abs/2502.18595},
	doi = {10.1007/JHEP08(2025)217},
	number = {8},
	urldate = {2026-07-20},
	journal = {Journal of High Energy Physics},
	author = {Zhang, Ziruo and Peng, Cheng},
	month = aug,
	year = {2025},
	note = {arXiv:2502.18595 [hep-th]},
	pages = {217},
}

@misc{akyuz_four-point_2026,
	title = {Four-point function of the complex {Sachdev}-{Ye}-{Kitaev} model at finite chemical potential},
	url = {http://arxiv.org/abs/2601.04330},
	doi = {10.48550/arXiv.2601.04330},
	urldate = {2026-07-20},
	publisher = {arXiv},
	author = {Akyuz, Can Onur and Cruz, Erick Arguello and Fraser-Taliente, Ludo and Tarnopolsky, Grigory},
	month = jan,
	year = {2026},
	note = {arXiv:2601.04330 [hep-th]},
}

@article{gubankova_scaling_2026,
	title = {Scaling limits of complex {Sachdev}-{Ye}-{Kitaev} models and holographic geometry},
	volume = {2026},
	issn = {1029-8479},
	url = {http://arxiv.org/abs/2512.05294},
	doi = {10.1007/JHEP03(2026)249},
	number = {3},
	urldate = {2026-07-20},
	journal = {Journal of High Energy Physics},
	author = {Gubankova, Elena and Sachdev, Subir and Tarnopolsky, Grigory},
	month = mar,
	year = {2026},
	note = {arXiv:2512.05294 [hep-th]},
	pages = {249},
}

@misc{pethybridge_notes_2024,
	title = {Notes on complex \$q=2\$ {SYK}},
	url = {http://arxiv.org/abs/2403.04673},
	doi = {10.48550/arXiv.2403.04673},
	urldate = {2026-07-20},
	publisher = {arXiv},
	author = {Pethybridge, Benjamin James},
	month = apr,
	year = {2024},
	note = {arXiv:2403.04673 [hep-th]},
}

@article{gu_notes_2020,
	title = {Notes on the complex {Sachdev}-{Ye}-{Kitaev} model},
	volume = {2020},
	issn = {1029-8479},
	url = {http://arxiv.org/abs/1910.14099},
	doi = {10.1007/JHEP02(2020)157},
	number = {2},
	urldate = {2026-07-20},
	journal = {Journal of High Energy Physics},
	author = {Gu, Yingfei and Kitaev, Alexei and Sachdev, Subir and Tarnopolsky, Grigory},
	month = feb,
	year = {2020},
	note = {arXiv:1910.14099 [hep-th]},
	pages = {157},
}

@article{tikhanovskaya_excitation_2021,
	title = {Excitation spectra of quantum matter without quasiparticles {II}: random \$t\$-\${J}\$ models},
	volume = {103},
	issn = {2469-9950, 2469-9969},
	shorttitle = {Excitation spectra of quantum matter without quasiparticles {II}},
	url = {http://arxiv.org/abs/2012.14449},
	doi = {10.1103/PhysRevB.103.075142},
	number = {7},
	urldate = {2026-07-20},
	journal = {Physical Review B},
	author = {Tikhanovskaya, Maria and Guo, Haoyu and Sachdev, Subir and Tarnopolsky, Grigory},
	month = feb,
	year = {2021},
	note = {arXiv:2012.14449 [cond-mat.str-el]},
	pages = {075142},
}

@article{berkooz_double_2020,
	title = {The double scaled limit of {Super}--{Symmetric} {SYK} models},
	volume = {2020},
	issn = {1029-8479},
	url = {http://arxiv.org/abs/2003.04405},
	doi = {10.1007/JHEP12(2020)110},
	number = {12},
	urldate = {2026-07-20},
	journal = {Journal of High Energy Physics},
	author = {Berkooz, Micha and Brukner, Nadav and Narovlansky, Vladimir and Raz, Amir},
	month = dec,
	year = {2020},
	note = {arXiv:2003.04405 [hep-th]},
	pages = {110},
}

@article{benini_mathcaln2_2025,
	title = {{$\mathcal{N}=2$} {SYK} models with dynamical bosons and fermions},
	volume = {112},
	issn = {2470-0010, 2470-0029},
	url = {http://arxiv.org/abs/2402.08414},
	doi = {10.1103/jky3-v5dz},
	number = {6},
	urldate = {2026-07-20},
	journal = {Physical Review D},
	author = {Benini, Francesco and Reis, Tomas and Soltani, Saman and Zhang, Ziruo},
	month = sep,
	year = {2025},
	note = {arXiv:2402.08414 [hep-th]},
	pages = {065010},
}

@article{murugan_more_2017,
	title = {More on {Supersymmetric} and 2d {Analogs} of the {SYK} {Model}},
	volume = {2017},
	issn = {1029-8479},
	url = {http://arxiv.org/abs/1706.05362},
	doi = {10.1007/JHEP08(2017)146},
	number = {8},
	urldate = {2026-07-20},
	journal = {Journal of High Energy Physics},
	author = {Murugan, Jeff and Stanford, Douglas and Witten, Edward},
	month = aug,
	year = {2017},
	note = {arXiv:1706.05362 [hep-th]},
	pages = {146},
}

@misc{heydeman_supersymmetry_2024,
	title = {Supersymmetry breaking in {SYK} and the black hole spectrum},
	url = {http://arxiv.org/abs/2408.12138},
	doi = {10.48550/arXiv.2408.12138},
	urldate = {2026-07-20},
	publisher = {arXiv},
	author = {Heydeman, Matthew and Liu, Yifei and Turiaci, Gustavo J.},
	month = aug,
	year = {2024},
	note = {arXiv:2408.12138 [hep-th]},
}

@article{marcus_new_2019,
	title = {A new class of {SYK}-like models with maximal chaos},
	volume = {2019},
	issn = {1029-8479},
	url = {http://arxiv.org/abs/1808.01190},
	doi = {10.1007/JHEP01(2019)166},
	number = {1},
	urldate = {2026-07-20},
	journal = {Journal of High Energy Physics},
	author = {Marcus, Eric and Vandoren, Stefan},
	month = jan,
	year = {2019},
	note = {arXiv:1808.01190 [hep-th]},
	pages = {166},
}

@article{peng_correlators_2017,
	title = {Correlators in the {$\mathcal{N}=2$} {Supersymmetric} {SYK} {Model}},
	volume = {2017},
	issn = {1029-8479},
	url = {http://arxiv.org/abs/1706.06078},
	doi = {10.1007/JHEP10(2017)202},
	number = {10},
	urldate = {2026-07-20},
	journal = {Journal of High Energy Physics},
	author = {Peng, Cheng and Spradlin, Marcus and Volovich, Anastasia},
	month = oct,
	year = {2017},
	note = {arXiv:1706.06078 [hep-th]},
	pages = {202},
}

@article{gates_1d_2021,
	title = {On {1D}, {N} = 4 {Supersymmetric} {SYK}-{Type} {Models} ({I})},
	volume = {2021},
	issn = {1029-8479},
	url = {http://arxiv.org/abs/2103.11899},
	doi = {10.1007/JHEP06(2021)158},
	number = {6},
	urldate = {2026-07-20},
	journal = {Journal of High Energy Physics},
	author = {Gates, S. James and Hu, Yangrui and Mak, S.-N. Hazel},
	month = jun,
	year = {2021},
	note = {arXiv:2103.11899 [hep-th]},
	pages = {158},
}

@article{heydeman_phases_2023,
	title = {Phases of {$\mathcal{N}=2$} {Sachdev}-{Ye}-{Kitaev} models},
	volume = {2023},
	issn = {1029-8479},
	url = {http://arxiv.org/abs/2206.14900},
	doi = {10.1007/JHEP01(2023)098},
	number = {1},
	urldate = {2026-07-20},
	journal = {Journal of High Energy Physics},
	author = {Heydeman, Matthew and Turiaci, Gustavo J. and Zhao, Wenli},
	month = jan,
	year = {2023},
	note = {arXiv:2206.14900 [hep-th]},
	pages = {98},
}

@article{biggs_supersymmetric_2024,
	title = {A supersymmetric {SYK} model with a curious low energy behavior},
	volume = {2024},
	issn = {1029-8479},
	url = {http://arxiv.org/abs/2309.08818},
	doi = {10.1007/JHEP08(2024)124},
	number = {8},
	urldate = {2026-07-20},
	journal = {Journal of High Energy Physics},
	author = {Biggs, Anna and Maldacena, Juan and Narovlansky, Vladimir},
	month = aug,
	year = {2024},
	note = {arXiv:2309.08818 [hep-th]},
	pages = {124},
}

@article{peng_soft_2021,
	title = {Soft modes in {$\mathcal{N}=2$} {SYK} model},
	volume = {2021},
	issn = {1029-8479},
	url = {http://arxiv.org/abs/2006.13961},
	doi = {10.1007/JHEP01(2021)082},
	number = {1},
	urldate = {2026-07-20},
	journal = {Journal of High Energy Physics},
	author = {Peng, Cheng and Stanojevic, Stefan},
	month = jan,
	year = {2021},
	note = {arXiv:2006.13961 [hep-th]},
	pages = {82},
}

@article{fu_supersymmetric_2017,
	title = {Supersymmetric {SYK} models},
	volume = {95},
	issn = {2470-0010, 2470-0029},
	url = {http://arxiv.org/abs/1610.08917},
	doi = {10.1103/PhysRevD.95.069904},
	number = {6},
	urldate = {2026-07-20},
	journal = {Physical Review D},
	author = {Fu, Wenbo and Gaiotto, Davide and Maldacena, Juan and Sachdev, Subir},
	month = mar,
	year = {2017},
	note = {arXiv:1610.08917 [hep-th]},
	pages = {069904},
}

@article{gindikin_collective_2025,
	title = {Collective excitations and stability of a non-{Fermi} liquid state near a quantum-critical point of a metal},
	volume = {112},
	issn = {2469-9950, 2469-9969},
	url = {http://arxiv.org/abs/2505.08925},
	doi = {10.1103/cbkc-gj3b},
	number = {8},
	urldate = {2026-07-20},
	journal = {Physical Review B},
	author = {Gindikin, Yasha and Maslov, Dmitrii L. and Chubukov, Andrey V.},
	month = aug,
	year = {2025},
	note = {arXiv:2505.08925 [cond-mat.str-el]},
	pages = {L081101},
}

@misc{abanov_non-bcs_2024,
	title = {Non-{BCS} behavior of the pairing susceptibility near the onset of superconductivity in a quantum-critical metal},
	url = {http://arxiv.org/abs/2412.03698},
	doi = {10.48550/arXiv.2412.03698},
	urldate = {2026-07-20},
	publisher = {arXiv},
	author = {Abanov, Artem and Zhang, Shang-Shun and Chubukov, Andrey},
	month = dec,
	year = {2024},
	note = {arXiv:2412.03698 [cond-mat.supr-con]},
}

@misc{liu_reply_2024,
	title = {Reply to ``{Comment} on `{Towards} exact solutions of superconducting {$T_c$} induced by electron-phonon interaction'{}''},
	url = {http://arxiv.org/abs/2408.03857},
	doi = {10.48550/arXiv.2408.03857},
	urldate = {2026-07-20},
	publisher = {arXiv},
	author = {Liu, Guo-Zhu and Yang, Zhao-Kun and Pan, Xiao-Yin and Wang, Jing-Rong and Li, Xin and Zhu, Hao-Fu and Huang, Jie},
	month = aug,
	year = {2024},
	note = {arXiv:2408.03857 [cond-mat.supr-con]},
}

@article{palle_comment_2024,
	title = {Comment on ``{Towards} exact solutions for the superconducting {$T_c$} induced by electron-phonon interaction''},
	volume = {110},
	issn = {2469-9950, 2469-9969},
	url = {http://arxiv.org/abs/2404.02918},
	doi = {10.1103/PhysRevB.110.026501},
	number = {2},
	urldate = {2026-07-20},
	journal = {Physical Review B},
	author = {Palle, Grgur},
	month = jul,
	year = {2024},
	note = {arXiv:2404.02918 [cond-mat.supr-con]},
	pages = {026501},
}

@article{liu_towards_2021,
	title = {Towards exact solutions for the superconducting {$T_c$} induced by electron-phonon interaction},
	volume = {103},
	issn = {2469-9950, 2469-9969},
	url = {http://arxiv.org/abs/1911.05528},
	doi = {10.1103/PhysRevB.103.094501},
	number = {9},
	urldate = {2026-07-20},
	journal = {Physical Review B},
	author = {Liu, Guo-Zhu and Yang, Zhao-Kun and Pan, Xiao-Yin and Wang, Jing-Rong},
	month = mar,
	year = {2021},
	note = {arXiv:1911.05528 [cond-mat.supr-con]},
	pages = {094501},
}

@article{pandey_ward_2024,
	title = {Ward identity preserving approach for investigation of phonon spectrum with self-energy and vertex corrections},
	volume = {110},
	issn = {2469-9950, 2469-9969},
	url = {http://arxiv.org/abs/2308.14379},
	doi = {10.1103/PhysRevB.110.064321},
	number = {6},
	urldate = {2026-07-20},
	journal = {Physical Review B},
	author = {Pandey, Sudhakar},
	month = aug,
	year = {2024},
	note = {arXiv:2308.14379 [cond-mat.str-el]},
	pages = {064321},
}

@book{ammon_gaugegravity_2015,
	edition = {1},
	title = {Gauge/{Gravity} {Duality}: {Foundations} and {Applications}},
	copyright = {https://www.cambridge.org/core/terms},
	isbn = {978-0-511-84637-3 978-1-107-01034-5},
	shorttitle = {Gauge/{Gravity} {Duality}},
	url = {https://www.cambridge.org/core/product/identifier/9780511846373/type/book},
	doi = {10.1017/CBO9780511846373},
	urldate = {2026-07-20},
	publisher = {Cambridge University Press},
	author = {Ammon, Martin and Erdmenger, Johanna},
	month = apr,
	year = {2015},
}

@book{zaanen_holographic_2015,
	title = {Holographic {Duality} in {Condensed} {Matter} {Physics}},
	isbn = {978-1-107-08008-9},
	doi = {10.1017/CBO9781139942492},
	publisher = {Cambridge Univ. Press},
	author = {Zaanen, Jan and Sun, Ya-Wen and Liu, Yan and Schalm, Koenraad},
	year = {2015},
}

@article{sachdev_what_2012,
	title = {What can gauge-gravity duality teach us about condensed matter physics?},
	volume = {3},
	issn = {1947-5454, 1947-5462},
	url = {http://arxiv.org/abs/1108.1197},
	doi = {10.1146/annurev-conmatphys-020911-125141},
	number = {1},
	urldate = {2026-07-20},
	journal = {Annual Review of Condensed Matter Physics},
	author = {Sachdev, Subir},
	month = mar,
	year = {2012},
	note = {arXiv:1108.1197 [cond-mat.str-el]},
	pages = {9--33},
}

@article{mcgreevy_holographic_2010,
	title = {Holographic duality with a view toward many-body physics},
	volume = {2010},
	issn = {1687-7357, 1687-7365},
	url = {http://arxiv.org/abs/0909.0518},
	doi = {10.1155/2010/723105},
	number = {1},
	urldate = {2026-07-20},
	journal = {Advances in High Energy Physics},
	author = {McGreevy, John},
	month = jan,
	year = {2010},
	note = {arXiv:0909.0518 [hep-th]},
	pages = {723105},
}

@article{hartnoll_lectures_2009,
	title = {Lectures on holographic methods for condensed matter physics},
	volume = {26},
	doi = {10.1088/0264-9381/26/22/224002},
	journal = {Class. Quant. Grav.},
	author = {Hartnoll, Sean A.},
	editor = {Uranga, A. M.},
	year = {2009},
	note = {arXiv:0903.3246},
	pages = {224002},
}

@article{jevicki_bi-local_2016,
	title = {Bi-{Local} {Holography} in the {SYK} {Model}},
	volume = {2016},
	issn = {1029-8479},
	url = {http://arxiv.org/abs/1603.06246},
	doi = {10.1007/JHEP07(2016)007},
	number = {7},
	urldate = {2026-07-20},
	journal = {Journal of High Energy Physics},
	author = {Jevicki, Antal and Suzuki, Kenta and Yoon, Junggi},
	month = jul,
	year = {2016},
	note = {arXiv:1603.06246 [hep-th]},
	pages = {7},
}

@article{jevicki_bi-local_2016-1,
	title = {Bi-local holography in the {SYK} model: perturbations},
	volume = {2016},
	issn = {1029-8479},
	shorttitle = {Bi-local holography in the {SYK} model},
	url = {https://doi.org/10.1007/JHEP11(2016)046},
	doi = {10.1007/JHEP11(2016)046},
	number = {11},
	urldate = {2026-07-20},
	journal = {Journal of High Energy Physics},
	author = {Jevicki, Antal and Suzuki, Kenta},
	month = nov,
	year = {2016},
	pages = {46},
}

@article{das_three_2017,
	title = {Three {Dimensional} {View} of the {SYK}/{AdS} {Duality}},
	volume = {2017},
	issn = {1029-8479},
	url = {http://arxiv.org/abs/1704.07208},
	doi = {10.1007/JHEP09(2017)017},
	number = {9},
	urldate = {2026-07-20},
	journal = {Journal of High Energy Physics},
	author = {Das, Sumit R. and Jevicki, Antal and Suzuki, Kenta},
	month = sep,
	year = {2017},
	note = {arXiv:1704.07208 [hep-th]},
	pages = {17},
}

@article{das_space-time_2018,
	title = {Space-time in the {SYK} model},
	volume = {2018},
	issn = {1029-8479},
	url = {https://doi.org/10.1007/JHEP07(2018)184},
	doi = {10.1007/JHEP07(2018)184},
	number = {7},
	urldate = {2026-07-20},
	journal = {Journal of High Energy Physics},
	author = {Das, Sumit R. and Ghosh, Animik and Jevicki, Antal and Suzuki, Kenta},
	month = jul,
	year = {2018},
	pages = {184},
}

@article{mandal_coadjoint_2017,
	title = {Coadjoint orbit action of {Virasoro} group and two-dimensional quantum gravity dual to {SYK}/tensor models},
	volume = {2017},
	issn = {1029-8479},
	url = {https://doi.org/10.1007/JHEP11(2017)046},
	doi = {10.1007/JHEP11(2017)046},
	number = {11},
	urldate = {2026-07-20},
	journal = {Journal of High Energy Physics},
	author = {Mandal, Gautam and Nayak, Pranjal and Wadia, Spenta R.},
	month = nov,
	year = {2017},
	pages = {46},
}

@article{gaikwad_holographic_2020,
	title = {Holographic dual to charged {SYK} from {3D} {Gravity} and {Chern}-{Simons}},
	volume = {2020},
	issn = {1029-8479},
	url = {http://arxiv.org/abs/1802.07746},
	doi = {10.1007/JHEP02(2020)033},
	number = {2},
	urldate = {2026-07-20},
	journal = {Journal of High Energy Physics},
	author = {Gaikwad, Adwait and Joshi, Lata Kh and Mandal, Gautam and Wadia, Spenta R.},
	month = feb,
	year = {2020},
	note = {arXiv:1802.07746 [hep-th]},
	pages = {33},
}

@article{mertens_defects_2019,
	title = {Defects in {Jackiw}-{Teitelboim} {Quantum} {Gravity}},
	volume = {2019},
	issn = {1029-8479},
	url = {http://arxiv.org/abs/1904.05228},
	doi = {10.1007/JHEP08(2019)127},
	number = {8},
	urldate = {2026-07-20},
	journal = {Journal of High Energy Physics},
	author = {Mertens, Thomas G. and Turiaci, Gustavo J.},
	month = aug,
	year = {2019},
	note = {arXiv:1904.05228 [hep-th]},
	pages = {127},
}

@article{khveshchenko_diagnostic_2025,
	title = {Diagnostic {Tomography} of {Applied} {Holography}},
	volume = {39},
	issn = {0217-9792, 1793-6578},
	url = {http://arxiv.org/abs/2310.02991},
	doi = {10.1142/S0217979225501498},
	number = {17},
	urldate = {2026-07-20},
	journal = {International Journal of Modern Physics B},
	author = {Khveshchenko, D. V.},
	month = jul,
	year = {2025},
	note = {arXiv:2310.02991 [cond-mat.str-el]},
	pages = {2550149},
}

@misc{khveshchenko_it_2023,
	title = {{IT} from {QUBIT} or {ALL} from {HALL}?},
	url = {http://arxiv.org/abs/2305.04399},
	doi = {10.48550/arXiv.2305.04399},
	urldate = {2026-07-20},
	publisher = {arXiv},
	author = {Khveshchenko, D. V.},
	month = may,
	year = {2023},
	note = {arXiv:2305.04399 [hep-th]},
}

@article{khveshchenko_phase_2022,
	title = {Phase space holography with no strings attached},
	volume = {61},
	issn = {2424-3647, 1648-8504},
	url = {http://arxiv.org/abs/2102.01617},
	doi = {10.3952/physics.v61i4.4642},
	number = {4},
	urldate = {2026-07-20},
	journal = {Lithuanian Journal of Physics},
	author = {Khveshchenko, D. V.},
	month = jan,
	year = {2022},
	note = {arXiv:2102.01617 [cond-mat.str-el]},
}

@article{davison_thermoelectric_2017,
	title = {Thermoelectric transport in disordered metals without quasiparticles: the {SYK} models and holography},
	volume = {95},
	issn = {2469-9950, 2469-9969},
	shorttitle = {Thermoelectric transport in disordered metals without quasiparticles},
	url = {http://arxiv.org/abs/1612.00849},
	doi = {10.1103/PhysRevB.95.155131},
	number = {15},
	urldate = {2026-07-20},
	journal = {Physical Review B},
	author = {Davison, Richard A. and Fu, Wenbo and Georges, Antoine and Gu, Yingfei and Jensen, Kristan and Sachdev, Subir},
	month = apr,
	year = {2017},
	note = {arXiv:1612.00849 [cond-mat.str-el]},
	pages = {155131},
}

@article{cheipesh_planckian_2019,
	title = {Planckian superconductor},
	volume = {100},
	issn = {2469-9950, 2469-9969},
	url = {http://arxiv.org/abs/1910.00671},
	doi = {10.1103/PhysRevB.100.220506},
	number = {22},
	urldate = {2026-07-20},
	journal = {Physical Review B},
	author = {Cheipesh, Y. and Pavlov, A. I. and Scopelliti, V. and Tworzydlo, J. and Gnezdilov, N. V.},
	month = dec,
	year = {2019},
	note = {arXiv:1910.00671 [cond-mat.str-el]},
	pages = {220506},
}

@article{inkof_quantum_2022,
	title = {Quantum critical {Eliashberg} theory, the {SYK} superconductor and their holographic duals},
	volume = {7},
	issn = {2397-4648},
	url = {http://arxiv.org/abs/2108.11392},
	doi = {10.1038/s41535-022-00460-8},
	number = {1},
	urldate = {2026-07-20},
	journal = {npj Quantum Materials},
	author = {Inkof, Gian Andrea and Schalm, Koenraad and Schmalian, Jörg},
	month = jun,
	year = {2022},
	note = {arXiv:2108.11392 [cond-mat.str-el]},
	pages = {56},
}

@misc{esterlis_quantum_2025,
	title = {Quantum {Critical} {Eliashberg} {Theory}},
	url = {http://arxiv.org/abs/2506.11952},
	doi = {10.48550/arXiv.2506.11952},
	urldate = {2026-07-20},
	publisher = {arXiv},
	author = {Esterlis, Ilya and Schmalian, Joerg},
	month = jun,
	year = {2025},
	note = {arXiv:2506.11952 [cond-mat.str-el]},
}

@misc{shankar_josephson_2025,
	title = {A {Josephson} wormhole in coupled superconducting {Yukawa}-{SYK} metals},
	url = {http://arxiv.org/abs/2506.16907},
	doi = {10.48550/arXiv.2506.16907},
	urldate = {2026-07-20},
	publisher = {arXiv},
	author = {Shankar, Aravindh S. and Steenbergen, Jasper and Plugge, Stephan and Schalm, Koenraad},
	month = jun,
	year = {2025},
	note = {arXiv:2506.16907 [cond-mat.str-el]},
}

@misc{stone_radon_2025,
	title = {Radon {Transforms} and the {SYK} model},
	url = {http://arxiv.org/abs/2506.09880},
	doi = {10.48550/arXiv.2506.09880},
	urldate = {2026-07-20},
	publisher = {arXiv},
	author = {Stone, Michael},
	month = jun,
	year = {2025},
	note = {arXiv:2506.09880 [math-ph]},
}

@misc{abchub,
  author        = {Abanov, Artem G. and Chubukov, Andrey V.},
  title         = {{Non-BCS} Pairing by a Singular Dynamical Interaction},
  year          = {2026},
  month         = jun,
  eprint        = {2606.21731},
  archivePrefix = {arXiv},
  primaryClass  = {cond-mat.str-el},
  doi           = {10.48550/arXiv.2606.21731},
  url           = {https://arxiv.org/abs/2606.21731}
}

@misc{schmalian_last,
  author        = {Stangier, Veronika Carolin and Schmalian, J{\"o}rg},
  title         = {Why {Cooper} pairs live in {AdS2}: a spectral analysis of the {Yukawa-SYK} model},
  year          = {2026},
  month         = aug,
  eprint        = {2608.26251},
  archivePrefix = {arXiv},
  primaryClass  = {cond-mat.str-el},
  doi           = {10.48550/arXiv.2608.26251},
  url           = {https://arxiv.org/abs/2608.26251}
}

\end{document}